\documentclass[a4paper,11pt]{article}
\pdfoutput=1 

\usepackage{jheppub} 
\usepackage{amsmath, amsfonts}

\usepackage[T1]{fontenc} 
\usepackage{xcolor}
\usepackage{subfigure}
\usepackage{stmaryrd}
\usepackage{enumerate}
\usepackage{tikz}
\usetikzlibrary{calc,intersections,positioning,arrows}
\usepackage[english]{babel}
\usepackage{graphicx}
\usepackage{bm}
\usepackage{epsf}

\newcommand{\beq}{\begin{equation}}
\newcommand{\eeq}{\end{equation}}
\newcommand{\bea}{\begin{eqnarray}}
\newcommand{\eea}{\end{eqnarray}}

\usepackage{color}

\usepackage[normalem]{ulem}  

\newcommand{\non}{\nonumber\\}

\newcommand{\be}{\begin{equation}}
\newcommand{\ee}{\end{equation}}

\title{\boldmath Chiral Soliton Lattice turns into 3D  crystal}

\author[a,b,c]{Geraint W.\ Evans}
\author[c]{and Andreas Schmitt}

\affiliation[a]{Institute of Physics, Academia Sinica, Academia Rd, Taipei 11529, Taiwan.}
\affiliation[b]{Institut f\"ur Theoretische Physik, Goethe University, Max-von-Laue-Straße 1, 60438 Frankfurt am Main, Germany.}
\affiliation[c]{Mathematical Sciences and STAG Research Centre, University of Southampton, Highfield Campus, Southampton
SO17 1BJ, United Kingdom.}

\emailAdd{geraint@gate.sinica.edu.tw}
\emailAdd{a.schmitt@soton.ac.uk}

\abstract{Chiral perturbation theory   predicts the chiral anomaly to induce a so-called Chiral Soliton Lattice at sufficiently large magnetic fields and baryon chemical potentials. This state breaks translational invariance in the direction of the magnetic field and was shown to be unstable with respect to charged pion condensation. Improving on previous work by considering a realistic pion mass, we employ methods from type-II superconductivity and construct a three-dimensional pion (and baryon) crystal perturbatively, close to the instability curve of the Chiral Soliton Lattice. We find an analogue of the usual type-I/type-II transition in superconductivity: Along the instability curve for magnetic fields $eB > 0.12\, {\rm GeV}^2$ and chemical potentials $\mu< 910\, {\rm MeV}$, this crystal can continuously supersede the Chiral Soliton Lattice. For smaller magnetic fields the instability curve must be preceded by a discontinuous transition.  }

\begin{document} 
\maketitle
\flushbottom

\section{Introduction and conclusion}
\label{sec:intro}

\subsection{Context}
\label{sec:context}

Type-II superconductors can admit magnetic flux in the form of a flux tube lattice. Magnetic flux tube lattices have been predicted a long time ago by Abrikosov within Ginzburg-Landau theory \cite{Abrikosov:1957classic} and are routinely observed in superconductors in the laboratory \cite{Essman:1967ExpTri,HaradatypeII}. They may also play an important role in high-energy systems governed by Quantum Chromodynamics (QCD). For instance, they have been suggested to exist in the interior of neutron stars in the form of a proton superconductor in nuclear matter \cite{baympethick,Haber:2017kth,Wood:2020czv} or a color superconductor in quark matter \cite{Iida:2002ev,Alford:2010qf,Haber:2017oqb}, and may be found in the QCD phase diagram at nonzero isospin chemical potential in the form of a charged pion condensate \cite{Adhikari:2015wva}. It was pointed out in our previous work \cite{Evans:2022hwr} that elements of type-II superconductivity also apply to charged pion condensation without isospin chemical potential, but in the presence of a baryon chemical potential, which couples to the magnetic field and the pions through the chiral anomaly. In that study, we restricted ourselves for simplicity to the limit of zero pion mass and found a two-dimensional Abrikosov lattice, as in ordinary superconductors. In this paper, we generalize this study to the case of a physical pion mass and construct the resulting three-dimensional lattice. 

\subsection{Background and main idea}
\label{sec:background}

\begin{figure}[t]
\begin{center}
\includegraphics[width=0.5\textwidth]{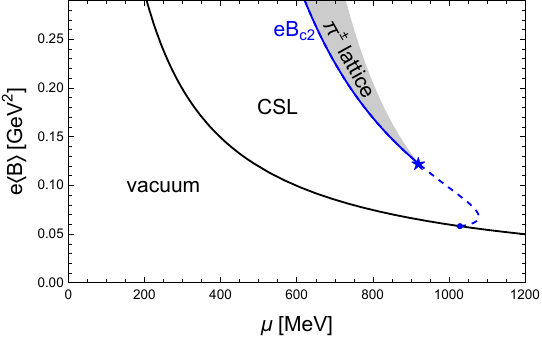}
\caption{Phase diagram in the plane of magnetic field and baryon chemical potential, showing our main results along the CSL instability curve $B_{c2}$ (blue). For large magnetic fields our perturbative analysis predicts a continuous transition to a three-dimensional charged pion lattice, indicated schematically by the shaded region. We find that this behavior changes at a point reminiscent of the type-I/type-II transition in ordinary superconductors (blue star): As one leaves the CSL region across the {\it dashed} segment of the $B_{c2}$ curve, our charged pion lattice is {\it not} the solution to the CSL instability. This includes the limit of a single domain wall (small blue dot at $e\langle B\rangle = 3m_\pi^2$). The diagram is obtained for realistic pion mass $m_\pi = 140\, {\rm MeV}$ and pion decay constant $f_\pi = 92\, {\rm MeV}$. In this case, the type-I/type-II transition point sits at $\mu\simeq 910\, {\rm MeV}$ and 
$e\langle B\rangle\simeq 0.12\, {\rm GeV}^2\simeq
6.3 \, m_\pi^2\simeq 6.1\times 10^{18}\, {\rm G}$. For decreasing (increasing) pion mass, this point moves towards larger (smaller) $\mu$. The result suggests additional phase transitions not captured by our approximation, see conjectured phase structure in Fig.\ \ref{fig:conjecture}.}
\label{fig:pd}
\end{center}
\end{figure}

The background for our study is the so-called Chiral Soliton Lattice (CSL). It was pointed out within chiral perturbation theory that the CSL phase is the ground state of QCD in a certain region of the phase diagram spanned by the magnetic field $B$ and the baryon chemical potential $\mu$ \cite{Brauner:2016pko} (see also studies in the chiral limit using the gauge-gravity duality \cite{Thompson:2008qw,Rebhan:2008ur,Preis:2011sp}). The crucial ingredient for this observation is the chiral anomaly \cite{Adler:1968anom,BellJackiw:1968anom},  implemented via a Wess-Zumino-Witten (WZW) term \cite{WessZumino:1971WZW,Witten:1983WZW}. This term results in an energy gain from a spatially varying neutral pion condensate. Fig.\ \ref{fig:pd} shows the phase transition line between the vacuum and CSL. At this phase transition it becomes energetically favorable to create a single domain wall with nonzero baryon number \cite{Son:2007ny}. Close to this transition line, CSL can be considered as a stack of well-separated domain walls. When they start to overlap they can no longer be described as independent domain walls -- the exact CSL solution needs to be considered. In the CSL phase, the magnetic field is homogeneous, say $\vec{B}=B \vec{e}_z$. In Fig.\ \ref{fig:pd}, we have generalized this to the average magnetic flux per area $\langle B\rangle$ through the transverse $x$-$y$ plane, kept constant along $z$. This allows for inhomogeneous magnetic fields and is necessary for our 3D lattice.  

The discussion of the CSL phase has been extended to nonzero temperatures \cite{Brauner:2021sci}, next-to-leading order contributions have been included \cite{Brauner:2023ort}, and its dynamical formation was studied \cite{Higaki:2022gnw}. Here we do not consider any of these extensions but focus on the instability of CSL with respect to charged pion condensation. This instability was already pointed out in the original work \cite{Brauner:2016pko}, without constructing the state that supersedes CSL beyond this instability. In Fig.\ \ref{fig:pd}, the corresponding instability curve (blue) is labelled by $B_{c2}$, borrowing the terminology from the second critical magnetic field in type-II superconductivity. In the chiral limit, i.e., setting the pion mass to zero, $m_\pi=0$, it was shown that the system forms magnetic flux tubes arranged in a hexagonal lattice in the transverse plane with respect to the magnetic field \cite{Evans:2022hwr}. This case is particularly simple because the system does not break translational invariance in the $z$-direction. Besides the magnetic field, also baryon number is concentrated in the flux tubes, which renders the system a 2D baryon crystal. The construction of this crystal was performed by solving the coupled equations of motion for the electromagnetic gauge field and the neutral and charged pion condensates in an expansion close to the critical field $B_{c2}$, following the Ginzburg-Landau analysis of type-II superconductivity \cite{Abrikosov:1957classic,Kleiner:1963Bulk,tinkham2004introduction}. 
In this paper, we shall use the same idea. However, the calculation is more involved because in the presence of a nonzero pion mass already the CSL phase itself breaks translational invariance in the direction of the magnetic field. Together with an Abrikosov lattice of charged pions in the transverse directions, the resulting phase can be expected to be a 3D crystal. Indeed, this is what we shall find: We will solve the (expanded) equations of motion in Fourier space to construct the crystal, where magnetic field, baryon number, and all pion condensates vary periodically in all three dimensions.

\subsection{Main results and discussion}
\label{sec:results}

The properties and precise structure of our 3D crystal will be discussed in the main part of the paper. Here we focus on the main results for the phase diagram.  

We find that our calculation provides us with a consistent solution only on one side of the instability curve $B_{c2}$, either outside or inside the CSL phase. If the solution exists inside CSL, we find that our 3D lattice is energetically disfavored compared to CSL, if it is outside we find that it is favored.
Importantly, there is a flip from outside (large $e\langle B\rangle$, solid segment of $B_{c2}$) to inside (small $e\langle B\rangle$, dashed segment of $B_{c2}$), marked by a star in Fig.\ \ref{fig:pd}. 
As a consequence, as one moves across the solid segment of the $B_{c2}$ line, a continuous transition from CSL to the 3D crystal is predicted, as indicated by the shaded region in the phase diagram. This is analogous to a transition from the normal-conducting state to the flux tube lattice in a type-II superconductor. In contrast, our lattice does not provide a solution for a continuous transition across the dashed segment. 

At the endpoint of the $B_{c2}$ curve (small blue dot) CSL is identical to a single domain wall. Therefore, our solution along $B_{c2}$ -- although not preferred everywhere -- interpolates from a 2D lattice with no dependence on $z$ (chiral limit of Ref.\ \cite{Evans:2022hwr}, valid for asymptotically large $B_{c2}$) to a lattice enclosed by a single domain wall localized in the $z$-direction. In particular, our perturbative  solution does not yield a stable lattice in the domain wall. 

The point marked by the star is reminiscent of the transition from type-II to type-I superconductivity. However, there is an important difference to ordinary superconductivity to be kept in mind. In our pionic system there is no equivalent of a Meissner phase. The would-be Meissner phase is a homogeneous charged pion condensate that completely expels the magnetic field. In the absence of a magnetic field, however, there is no CSL phase and thus no neutral pion condensate. Therefore, since we consider zero isospin chemical potential, there is no effective potential for the charged pions to condense in the first place.

\begin{figure}[t]
\begin{center}
\includegraphics[width=0.5\textwidth]{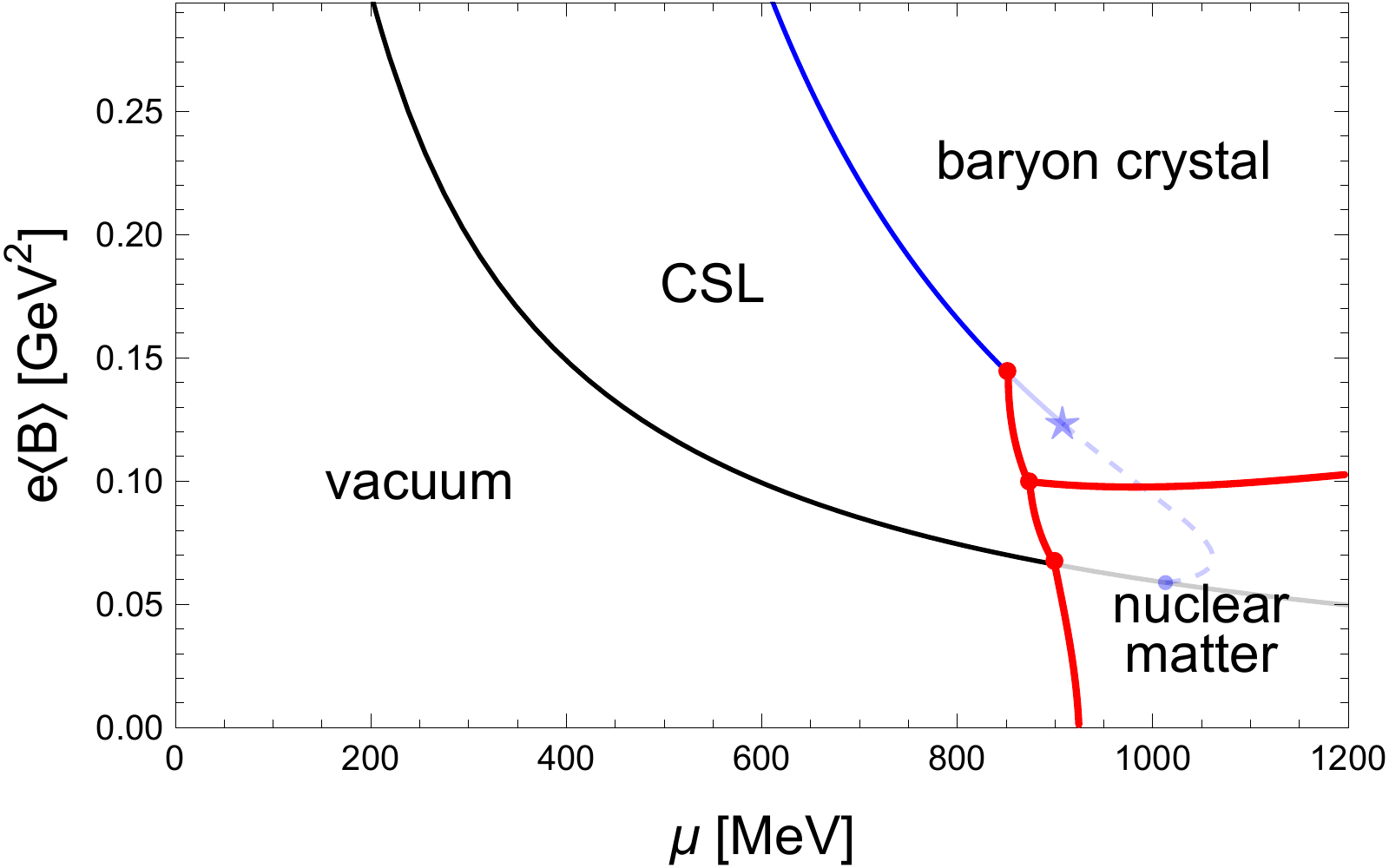}
\caption{Conjectured phase diagram, based on the results of Fig.\ \ref{fig:pd} plus the fact that at zero magnetic field there is a transition from the vacuum to homogeneous nuclear matter at $\mu_0\simeq 922.7\, {\rm MeV}$. Our type-I/type-II transition point (pale blue star) sets a limit for where the (red) discontinuous transition can attach to the (blue) continuous transition between CSL and the 3D baryon crystal. Pale segments are taken over from Fig.\ \ref{fig:pd} but do not correspond to actual transitions in our conjecture.}
\label{fig:conjecture}
\end{center}
\end{figure}

We cannot rigorously claim to have found the ground state of chiral perturbation theory; let alone the ground state of QCD, due to the limitations of chiral perturbation theory. Nevertheless,
it is tempting to speculate what our findings suggest for the QCD phase structure. In Fig.\ \ref{fig:conjecture} we present a phase diagram with conjectured phase transition lines (red), based on the following observations. Firstly, if one reaches the dashed segment of $B_{c2}$ from within CSL, one cannot transition to our charged pion lattice and yet we know that CSL becomes unstable there. Therefore, there must be an earlier transition either to the pion lattice or some other phase not considered in our calculation. This transition is necessarily discontinuous since it is not indicated by a CSL instability. It can connect to the continuous transition at any point on the solid $B_{c2}$ segment. Secondly, we know that at zero magnetic field there is a first-order transition from the vacuum to homogeneous nuclear matter at $\mu_0\simeq 922.7\, {\rm MeV}$\footnote{Interestingly, by only slightly varying the pion mass and the pion decay constant, using $m_\pi = 139\, {\rm MeV}$ and $f_\pi = 92.4\, {\rm MeV}$, we find that the chemical potential for our type-I/type-II transition point is $\mu\simeq 922\, {\rm MeV}$, strikingly close to $\mu_0$. However, there seems to be no reason why our transition point should reproduce the baryon onset $\mu_0$, mainly because it is located at a nonzero magnetic field, where the baryon onset is likely to be different from $\mu_0$ \cite{Rebhan:2008ur,Haber:2014ula}. Also, by decreasing $m_\pi$ at fixed $f_\pi$, our transition point moves towards larger chemical potentials. Since we do not expect the baryon mass to increase accordingly \cite{Procura:2003ig}, the transition point does not seem to be coupled to the baryon onset for all pion masses.}. Consequently, there must be at least one more transition, namely from homogeneous nuclear matter to the pion (and baryon) crystal. While the existence of such a liquid-solid transition seems inevitable, its exact location in our sketch is purely speculative. Our conjecture shows the simplest possible scenario to accommodate our findings with the existence of homogeneous nuclear matter at zero magnetic field. The phase structure may, however, be more complicated. For instance, we have ignored superfluidity and superconductivity in homogeneous nuclear matter, which may break down as the magnetic field is increased or which may turn via a single phase transition into the superconducting baryon crystal.
(The critical magnetic field for neutron and proton Cooper pairing depends on density with maximal values of the order of $10^{17} \, {\rm G}  \sim m_\pi^2$ \cite{Sinha:2015bva,Stein:2015bpa}.)

\subsection{Outlook}
\label{sec:outlook}

To put our results into context we should first recall that they are obtained within leading-order chiral perturbation theory. Going beyond leading order requires the calculation of the excitations of our lattice, which would be an interesting problem for future studies. In the regime where our calculation predicts a 3D pion crystal, the  energy scales set by the magnetic field and/or the chemical potential become comparable to $4\pi f_\pi$ and thus are sufficiently large for chiral perturbation theory itself to  approach its limit of validity. Hence we do not interpret our results as rigorous QCD predictions; ultimately, it would be interesting to compare our predictions with a first-principle QCD calculation. However, due to the presence of the chemical potential this seems to be out of reach for current methods in lattice gauge theory. Therefore, it would also be interesting to search for our 3D crystal within suitable models, such as nucleon-meson models or employing the gauge-gravity duality.

To explore our conjectured discontinuous phase transition lines, it would be necessary to go beyond our perturbative approach, which is only valid close to a continuous transition. Possibly this can be done  along the lines of previous studies \cite{Canfora:2020kyj,Canfora:2020uwf,Chen:2021vou,Canfora:2022jmh,Eto:2023lyo,Eto:2023wul}, which have constructed baryon crystals without using an expansion at the CSL instability, but which have made other simplifications compared to our study.
Reference \cite{Eto:2023wul}, which we learned of after completing the work reported here, predicts a discontinuous transition from the CSL phase to a phase made of domain-wall Skyrmion chains, however without allowing the magnetic field to be dynamical and without interaction of the chains in the transverse plane. It would be interesting to compare and combine these results with ours  -- which are fully dynamical in longitudinal and transverse directions, including the magnetic field, but only valid at a continuous transition. 

Since there exist various extensions and generalizations of the CSL phase, one can also think about applying these extensions to our crystal, for instance including a nonzero temperature \cite{Brauner:2021sci} or the $\eta$ meson \cite{Qiu:2023guy}, or replacing the magnetic field with an externally imposed rotation \cite{Huang:2017pqe,Eto:2021gyy,Eto:2023tuu}. Especially going beyond zero temperature might be crucial to connect our results to real-world systems. Large magnetic fields are reached in neutron stars and mergers thereof, as well as in heavy-ion collisions. In neutron star mergers and heavy-ion collisions, however, the zero-temperature approximation is not applicable. Also the magnetic field strengths where our crystal is stable are most likely pushing the limits of these systems. Therefore, it is of particular interest to find the critical magnetic field for our conjectured transition of liquid nuclear matter to the baryon crystal. 

Including an isospin chemical potential is another such extension. An isospin chemical potential gives rise to charged pion condensation more directly, without the detour via the chiral anomaly and neutral pion condensation. In the absence of a baryon chemical potential, a standard 2D vortex lattice can thus be induced by a magnetic field \cite{Adhikari:2015wva,Adhikari:2015pva}. It was later pointed out that the chiral anomaly allows for  a CSL phase in this case as well, and its competition with the vortex lattice was studied \cite{Gronli:2022cri} (however, without  taking into account  the possibility of a 3D lattice of coexisting neutral and charged pion condensates). 
It would be interesting to connect those findings to our results, i.e., to ask whether and how our anomaly-induced superconductivity relates to the more conventional superconductivity already present there. As a first step, this can be done by switching on an isospin chemical potential in our perturbative solution of the 3D crystal, or by searching for an instability of the isospin analogue of the CSL phase towards charged pion condensation.

\subsection{Structure of the paper}

The preparatory sections \ref{sec:lag} and \ref{sec:csl} serve to introduce our setup within chiral perturbation theory (Sec.\ \ref{sec:lag}) and to discuss the CSL phase and its instability (Sec.\ \ref{sec:csl}), thereby establishing our notation and collecting the necessary equations for the subsequent sections. Our main calculation, leading to the main results discussed in Sec.\ \ref{sec:results}, is laid out in Secs.\ \ref{sec:calc} and \ref{sec:free}. This contains the explanation of our expansion in Sec.\ \ref{sec:exp} and the construction and discussion of the 3D lattice in Secs.\ \ref{sec:phi0} -- \ref{sec:baryon}, as well as the calculation of the free energy in Secs.\ \ref{sec:general} -- \ref{sec:df}. The results that feed directly into the phase diagram in Fig.\ \ref{fig:pd} can be found at the end of Sec.\ \ref{sec:df}. 
We work in natural units $\hbar=c=k_B=1$ and use Heaviside-Lorentz units for the electromagnetic gauge field, in which the elementary charge is $e\simeq 0.30282$.

\section{Lagrangian}
\label{sec:lag}

 In this section, we write down our Lagrangian and collect the most important definitions. Since the starting point and notation are the same as in Ref.\ \cite{Evans:2022hwr}, we refer the reader to this reference for additional details. 
 We work within leading-order, two-flavor chiral perturbation theory with a dynamical electromagnetic field and a WZW term, such that the Lagrangian can be written as
\bea \label{Lfull}
{\cal L} = {\cal L}_{\rm em} + {\cal L}_\Sigma+ \mathcal{L}_{\text{WZW}} \, .
\eea
The electromagnetic part is 
\be \label{Lem}
{\cal L}_{\rm em} = -\frac{1}{4}F_{\mu\nu}F^{\mu\nu} \, , 
\ee
where $F^{\mu\nu} = \partial^\mu A^\nu - \partial^\nu A^\mu$ is the electromagnetic field strength tensor, with the electromagnetic $U(1)$ gauge field  $A^\mu$. Following the parameterization of Ref.\ \cite{Evans:2022hwr} (see also Ref.\ \cite{Brauner:2016pko}), we write the pionic contribution as 
\bea  \label{LSig2}
{\cal L}_\Sigma &=&
D_\mu \varphi(D^\mu\varphi)^* + \frac{\partial_\mu|\varphi|^2\partial^\mu|\varphi|^2}{2(f_\pi^2-2|\varphi|^2)}  + \frac{f_\pi^2-2|\varphi|^2}{2}\partial_\mu\alpha\partial^\mu\alpha\non[2ex]
&& +\,m_\pi^2f_\pi\sqrt{f_\pi^2-2|\varphi|^2}\cos\alpha \, , 
\eea
where 
\be
D^\mu\varphi = \partial^\mu\varphi+i(eA^\mu-\partial^\mu\alpha)\varphi
\ee
is the covariant derivative. The real scalar field $\alpha$ and the complex scalar field $\varphi$ parameterize the neutral and charged pions in a convenient way. In the most conventional parameterization, the chiral $SU(2)$ field is written as $\Sigma = (\sigma + i\pi_a\tau_a)/f_\pi$, where $a=1,2,3$, and $\tau_a$ are the Pauli matrices. In the $(\sigma,\pi_3)$ sector we introduce $\alpha$ as the polar angle and eliminate the modulus with the help of the constraint $f_\pi^2 = \sigma^2 + \pi_a\pi_a$. In the $(\pi_1,\pi_2)$ sector the complex field $\varphi$ is obtained from a rotation by the angle $\alpha$ of the field $(\pi_1+i\pi_2)/\sqrt{2}$. The WZW term can be written as
\begin{equation}
    \mathcal{L}_{\text{WZW}} = \left( A^B_{\mu} - \frac{e}{2}A_{\mu}\right)j^{\mu}_B\,,
    \label{ActionWZW}
\end{equation}
where $A_\mu^B = (\mu,0,0,0)$ contains the baryon chemical potential $\mu$, and the Goldstone-Wilczek baryon current \cite{Goldstone:1981kk,Son:2007ny,Brauner:2016pko} can be written as \cite{Evans:2022hwr}
\bea
    j^{\mu}_{B} &=&  -\frac{\epsilon^{\mu\nu\rho\lambda}}{4\pi^2}\partial_{\nu}\alpha
   \left(\frac{e}{2}F_{\rho\lambda}+ \frac{\partial_{\rho}j_{\lambda}}{ef_{\pi}^2}
        \right)\,,
    \label{BCurrent}
\eea  
with the non-anomalous contribution to the charge current 
\be
    j^{\mu} = ie\left(\varphi^*\partial^{\mu}\varphi - \varphi\partial^{\mu}\varphi^*\right) -2e\left(eA^{\mu}-\partial^{\mu}\alpha\right)|\varphi|^2 \, .
    \label{QCurrent}
\ee
Since we will only consider equilibrium configurations, we ignore any time dependence. Also, we only consider a magnetic field, $\vec{B} = \nabla\times\vec{A}$, neglecting any electric field contribution.  This requires the assumption that the system is locally neutral due to the addition of electrons or positrons that will not appear explicitly in our calculation (analogous to the usual Ginzburg-Landau study of superconductivity). We also perform our analysis on the classical level without taking into account fluctuations of the pions and the photon. Then, the local free energy density is simply given by $\Omega = -{\cal L}$ and can be written as    
\bea \label{Omega0}
    \Omega(\vec{x}) &=& \frac{B^2}{2}+\left|\left[\nabla -i\left(e\vec{ A}+\nabla\alpha\right)\right]\varphi\right|^2 +\frac{\left(\nabla|\varphi|^2\right)^2}{2\left(f_{\pi}^2-2|\varphi|^2\right)} +\frac{f_{\pi}^2 -2|\varphi|^2}{2}\left(\nabla\alpha\right)^2
        \non[2ex]
        &&-\,m_{\pi}^2f_{\pi}\sqrt{f_{\pi}^2-2|\varphi|^2}\cos{\alpha}
        -\mu n_B(\vec{x})         \, , 
\eea
where 
\begin{equation}
   n_{B}(\vec{x})= j_B^0 = \frac{\nabla\alpha}{4\pi^2}\cdot\left(e\vec{B}+\frac{\nabla\times\vec{j}}{ef_\pi^2}\right) 
   \label{nblocal}
\end{equation}
is the local baryon number density. The free energy density (\ref{Omega0}) can be viewed as an effective potential  and we can use it to derive the equations of motion for the dynamical fields $\vec{A}$, $\alpha$, $\varphi$. 

\section{Chiral Soliton Lattice and its instability}
\label{sec:csl}

Next, let us recapitulate the CSL phase and its instability \cite{Brauner:2016pko} to set the stage for our construction of the 3D pion lattice. 

\subsection{CSL solution}
\label{sec:csl1}

In the absence of charged pions, the effective potential is 
\bea \label{OmegaCSL}
\Omega_0(\vec{x})  =  \frac{B^2}{2}+\frac{f_\pi^2}{2}[(\nabla\alpha_0)^2-2m_\pi^2(\cos\alpha_0-1)]-\frac{e\mu}{4\pi^2}\nabla\alpha_0\cdot\vec{B} \, , 
\eea
where we have subtracted the constant $B=\alpha=0$ contribution to normalize the vacuum free energy to zero. Without charged pions, $\alpha$ is identical to the neutral pion field $\pi_3$, and we use the notation $\alpha_0$ to distinguish the CSL solution from our full solution in the subsequent sections. The equation of motion for $\alpha_0$ is 
\be 
\Delta\alpha_0 = m_\pi^2\sin\alpha_0 \, .
\ee
Aligning the $z$ axis with the magnetic field, $\vec{B}=B\vec{e}_z$, the physically relevant solution only depends on $z$. We write the solution as 
\be \label{cossn}
\alpha_0(\bar{z},p) = 2\,{\rm arccos}[-{\rm sn}(\bar{z},p^2)] \, , 
\ee
where sn denotes the Jacobi elliptic function, and $p\in [0,1]$ is a dimensionless number (the ``elliptic modulus'') whose value has to be determined by minimizing the free energy for given $\mu$ and $B$.  Moreover, we have introduced the dimensionless variable
\be
\bar{z} = \frac{zm_\pi}{p} \, .
\ee
The boundary conditions for the solution (\ref{cossn}) are chosen such that $\alpha_0$ winds from 0 to $2\pi$ in the unit cell $\bar{z}\in [-K(p^2),K(p^2)]$, where $K$ is the complete elliptic integral of the first kind. For $p\to 1$, the unit cell becomes infinitely large; this limit corresponds to an isolated domain wall, where the phase $\alpha_0$ winds around once in a spatial domain of order $m_\pi^{-1}$, 
\be \label{domain}
\mbox{domain wall limit:} \qquad \alpha_0(z) = 4\,{\rm arctan}\, e^{m_\pi z} \, .
\ee  
In the following we denote spatial averaging over the $z$ direction of any field $f(\vec{r})$  by 
\be \label{zAv}
\langle f(\vec{r}) \rangle_z= \frac{1}{2K(p^2)} \int_{-K(p^2)}^{K(p^2)} d\bar{z}\, f(\vec{r}) \, .
\ee
Since the CSL configuration is constant in the transverse directions, the average free energy density is given by 
\be \label{F0}
F_0 \equiv  \langle \Omega_0 \rangle_z  = \frac{B^2}{2} + 2m_\pi^2f_\pi^2 \left[1-\frac{1}{p^2}+\frac{2E(p^2)}{p^2K(p^2)}\right]-\frac{e\mu Bm_\pi}{4\pi p K(p^2)} \, ,
\ee
where $E$ is the complete elliptic integral of the second kind. 
Minimizing  $F_0$ with respect to $p$ gives
\be
\frac{E(p^2)}{p}=  \frac{eB\mu}{16\pi m_\pi f_\pi^2} \, , \label{minik}
\ee
which is an implicit equation for the elliptic modulus $p$. This yields the free energy at the minimum,
\be \label{F0min}
F_0 = \frac{B^2}{2} -2m_\pi^2f_\pi^2\left(\frac{1}{p^2}-1\right) \, .
\ee
We can use Eqs.\ (\ref{minik}) and (\ref{F0min}) to obtain a simple expression for the average baryon number density,
\be \label{nBCSL}
\langle n_B \rangle_z = -\frac{\partial F_0}{\partial\mu} = \frac{eB m_\pi}{4\pi pK(p^2)} \, .
\ee
We see from Eq.\ (\ref{F0min}) that the CSL state is favored over the vacuum for $p<1$ and the continuous transition occurs at $p=1$, the domain wall limit. With Eq.\ (\ref{minik}) and $E(1)=1$, the critical magnetic field for the onset of CSL is 
\be
eB_{\rm CSL}\equiv \frac{16\pi m_\pi f_\pi^2}{\mu} \, ,
\ee
given by the black curve in Fig.\ \ref{fig:pd}. We recover the chiral limit with $m_\pi\to 0$, $p\to 0$ and, using Eq.\ (\ref{minik}) with $E(0)=\pi/2$, 
\be\label{chiralmp}
\frac{m_\pi}{p} \to \frac{eB\mu}{8\pi^2f_\pi^2} \, .  
\ee
In this limit, the CSL solution (\ref{cossn}) reduces to 
\be \label{chiral}
\mbox{chiral limit:} \qquad \alpha_0(z) = \pi + \frac{eB\mu}{4\pi^2f_\pi^2}z \, .
\ee
Up to the irrelevant constant shift by $\pi$, this is the $m_\pi=0$ solution used in Ref.\ \cite{Evans:2022hwr}. While in Ref.\ \cite{Evans:2022hwr} the chiral limit was employed for all magnetic fields for simplicity, it is a valid approximation for magnetic fields much larger than the pion mass squared. Our full solution will also, just like $\alpha_0$ in this section, interpolate between the domain wall limit and the chiral limit.

\subsection{CSL instability}
\label{sec:csl2}

The CSL phase is unstable with respect to charged pion condensation \cite{Brauner:2016pko}. To observe this instability we go back to the original Lagrangian (\ref{Lfull}), including the complex field $\varphi$. The equation of motion for $\varphi^*$ is 
\bea \label{EOMphistar}
0&=& \left[D_{\mu}D^{\mu} +\partial_{\mu}\alpha\partial^{\mu}\alpha +\frac{\partial_{\mu}\partial^{\mu}|\varphi|^2}{f_{\pi}^2-2|\varphi|^2} +\frac{\partial_{\mu}|\varphi|^2\partial^{\mu}|\varphi|^2}{\left(f_{\pi}^2-2|\varphi|^2\right)^{2}}
+\frac{m_{\pi}^2 f_{\pi}\cos{\alpha}}{\sqrt{f_{\pi}^2 -2|\varphi|^2}}\right.\non[2ex]
&&
\left.+\frac{ie\epsilon^{\mu\nu\rho\lambda}}{8\pi^2f_{\pi}^2}\partial_{\nu}\alpha\, F_{\rho\lambda}D_{\mu}\right]\varphi\,.
\label{phiEOMFull}
\eea
We insert the CSL solution $\alpha_0$ into this equation and linearize in the field $\varphi$. Then, within our assumptions of a locally charge neutral system, setting $A_0=0$ and $e\vec{A}+\nabla\alpha_0=eBx\vec{e}_y$ and using the ansatz $\varphi(\vec{r},t) = e^{-i\omega t}e^{ik_y y}h(x,z)$, with the yet to be determined function $h(x,z)$, Eq.\ (\ref{EOMphistar}) becomes 
\be\label{omfxz}
\omega^2h = \left[-\partial_x^2+e^2B^2\left(x-\frac{k_y}{eB}\right)^2\right]h-\left\{\partial_z^2+\frac{m_\pi^2}{p^2} \left[4+p^2-6p^2{\rm sn}^2(\bar{z},p^2)\right]\right\}h \, ,
\ee
where we have used 
\be \label{dza0}
(\partial_z\alpha_0)^2-m_\pi^2\cos\alpha_0 = \frac{m_\pi^2}{p^2} \left[4+p^2-6p^2{\rm sn}^2(\bar{z},p^2)\right] \, .
\ee
We factorize  $h(x,z)= f(z)g(x)$ and divide by $f(z)$ to obtain an equation for $g(x)$ that has the form of the Schr\"{o}dinger equation for the harmonic oscillator. We can thus immediately read off the ($z$-dependent) energy eigenvalues labeled by $l = 0,1,2,\ldots$,  
\be 
\omega^2 = (2l+1)eB-\frac{m_\pi^2}{p^2}\left[4+p^2-6p^2{\rm sn}^2(\bar{z},p^2)\right]-\frac{f''}{f} \, .
\ee
Rearranging this equation gives
\be
Lf(\bar{z}) = \varepsilon f(\bar{z}) \, , 
\ee
with the Lam\'{e} operator
\be
L = -\partial_{\bar{z}}^2+m(m+1)p^2{\rm sn}^2(\bar{z},p^2) \, , 
\ee
where, in our case, $m=2$, and 
\be \label{calE}
\varepsilon = \frac{p^2}{m_\pi^2}[\omega^2-(2l+1)eB]+4+p^2
\, . 
\ee
The lowest eigenvalue of the $m=2$ Lam\'{e} equation 
is \cite{Finkel:1999Lame}
\be
\varepsilon  = 2(1+p^2-\sqrt{p^4-p^2+1}) \, , 
\ee
with corresponding eigenfunction 
\be \label{f0}
f_0(\bar{z}) = \frac{1}{\sqrt{N(p)}}\left(\frac{\sqrt{p^4-p^2+1}+1-2p^2}{3p^2}+\sin^2\frac{\alpha_0}{2}\right) \, ,
\ee
where
\be \label{Nnorm}
N(p)\equiv \frac{2\sqrt{p^4-p^2+1}}{9p^4}\left(\sqrt{p^4-p^2+1} +\frac{3E(p^2)}{K(p^2)}-2+p^2\right) \, .
\ee
This factor is introduced to ensure that $f_0$ is normalized, 
\be \label{normalize}
\langle f_0^2\rangle_z = 1 \, ,
\ee
which is convenient, although not strictly necessary, for our calculation. 
With Eq.\ (\ref{calE}) and setting $l=0$, the lowest-energy excitation is 
\be\label{omeBm}
\omega^2 = eB+m_\pi^2\left(1-2\frac{1+\sqrt{p^4-p^2+1}}{p^2}\right) \, .
\ee
The instability with respect to pion condensation sets in at $\omega^2=0$, i.e., at a critical field given by 
\be\label{nudef}
b \equiv \frac{eB_{c2}}{m_\pi^2} = \frac{2-p^2+2\sqrt{p^4-p^2+1}}{p^2} \, .
\ee
Our notation $B_{c2}$ anticipates that this critical field is reminiscent of the second critical magnetic field of a type-II superconductor. For the following it is useful to keep in mind that Eq.\ (\ref{nudef}) gives us a one-to-one relation between the critical magnetic field and the elliptic modulus $p\in [0,1]$. The CSL instability line is parameterized by $p$, from $p=0$, corresponding to infinitely large $b$, to $p=1$, corresponding to the end of the instability line at $b=3$, where CSL has turned into a single domain wall.
The corresponding chemical potential at the onset of the instability is then obtained from Eq.\ (\ref{minik}) as a function of $b$ (or $p$), 
\be \label{munu}
\mu = \frac{8\pi f_\pi^2}{m_\pi} \frac{\sqrt{(b+1)^2-4}}{b^{3/2}}E\left[\frac{4b}{(b+1)^2-4}\right] \, .
\ee
Again, it is useful to check the domain wall and chiral limits. For the domain wall, $p=1$, we easily obtain from Eq.\ (\ref{nudef}), 
\be \label{Bc2domain}
\mbox{domain wall:} \qquad eB_{c2}=3m_\pi^2 \, , 
\ee
in accordance with Ref.\ \cite{Son:2007ny}. In the chiral limit, $p\to 0$, we find with Eq.\ (\ref{chiralmp}), 
\be \label{Bc2chiral}
\mbox{chiral limit:} \qquad eB_{c2} = \frac{16\pi^4f_\pi^4}{\mu^2} \, , 
\ee
in accordance with Refs.\ \cite{Evans:2022hwr,Brauner:2016pko}. 

The instability curve given by Eq.\ (\ref{munu}) is shown as the blue line (solid and dashed segments together) in Fig.\ \ref{fig:pd} for the physical values of $m_\pi$ and $f_\pi$. The solution to the instability, at least close to the instability curve, is the main topic of this paper, to which we turn now. 

\section{Calculation of the 3D pion crystal}
\label{sec:calc}

\subsection{Expansion at the critical magnetic field}
\label{sec:exp}

At the second critical magnetic field of a type-II superconductor, there is a continuous transition from the normal-conducting state, where the magnetic field penetrates homogeneously,  to the superconducting state, where the magnetic field is confined into flux tubes. Magnetic flux tubes arise from a periodically varying charged condensate, which is zero along  the central axes of the flux tubes. At the transition, while the lattice constant of the flux tube lattice is nonzero and finite, the amplitude of the condensate is infinitesimally small. This allows for a small-field expansion, which we will also employ at our critical field $B_{c2}$. For a consistent power counting, let us introduce the smallness parameter 
\be \label{epsilon} 
\epsilon = \sqrt{\frac{|\langle B\rangle -B_{c2}|}{B_{c2}}} \, .
\ee
Here, as already introduced in the phase diagram in Fig.\ \ref{fig:pd}, $\langle B\rangle$ denotes the average magnetic flux per area through any $z={\rm const}$ plane. We will require this average flux to be the same at each $z$ and use $\langle B\rangle$ as our externally given thermodynamic variable\footnote{This is a slight abuse of notation because $\langle - \rangle$ will later be used for the spatial average across all three directions. Nevertheless, to connect to the notation of Ref.\ \cite{Evans:2022hwr} and to have a compact symbol for our thermodynamic variable we will keep using $\langle B\rangle \equiv \langle B_z\rangle_{x,y}$, where $\langle -\rangle_{x,y}$ means averaging over the transverse plane.}. For the CSL phase of the previous section, $\langle B\rangle = B$. We expand in $\epsilon$ at fixed baryon chemical potential $\mu$. Therefore, due to the two-valuedness of $B_{c2}$ as a function of $\mu$ in a certain regime (see Fig.\ \ref{fig:pd}) the region where we expect an Abrikosov lattice can either be at $\langle B\rangle > B_{c2}$ or at $\langle B\rangle < B_{c2}$, hence the modulus in Eq.\ (\ref{epsilon}). 
We denote the expansion of our dynamical fields by 
\be
\varphi = \varphi_0 + \delta \varphi \, , \qquad \vec{A} = \vec{A}_0 + \delta \vec{A} \, , \qquad \alpha = \alpha_0 +\delta \alpha \, , \ee
where $\varphi_0 \sim  \epsilon$, $\delta \varphi \sim \epsilon^3$, $\vec{A}_0\sim 1$, $\delta\vec{A}_0 \sim \epsilon^2$, $\alpha_0\sim 1$, $\delta\alpha\sim\epsilon^2$. The expansion of the gauge field implies for the magnetic field 
\be \label{BB0}
\vec{B} = \vec{B}_0 + \delta \vec{B} \, , 
\ee
with $\vec{B}_0 = B_{c2}\vec{e}_z$ of order 1 and $\delta\vec{B}\sim \epsilon^2$. We can now write down the equations of motion derived from the effective potential (\ref{Omega0}) order by order in $\epsilon$. The leading and next-to-leading order equations for each field are \cite{Evans:2022hwr}
\begin{subequations}\label{EoMEx} \allowdisplaybreaks
\bea
{\cal D}_0\varphi_0&=&0  \,, \label{phi0EoMEx} \\[2ex]
{\cal D}_0\delta\varphi &=& \Bigg[2i(e\delta\vec{A}+\nabla\delta\alpha)\cdot\nabla+i\nabla\cdot(e\delta\vec{A}+\nabla\delta\alpha)+2(e\vec{A}_0+\nabla\alpha_0)\cdot(e\delta\vec{A}+\nabla\delta\alpha)\non[2ex]
&&-(\Delta\alpha_0+2\nabla\alpha_0\cdot\nabla)\delta\alpha-\frac{(\Delta-m_\pi^2\cos\alpha_0)|\varphi_0|^2}{f_\pi^2}\Bigg]\varphi_0 \,  ,\label{deltaphiEoMEx}\\[2ex]
\Delta\alpha_0 &=& m_\pi^2 \sin\alpha_0 \, , \label{alpha0EoMEx} \\[2ex]
0&=&f_\pi^2(\Delta-m_\pi^2\cos\alpha_0)\delta\alpha -(\Delta\alpha_0+2\nabla\alpha_0\cdot\nabla) |\varphi_0|^2  \, , \label{deltaalphaEoMEx}\\[2ex]
\nabla\times \vec{B}_0 &=& 0 \, , \label{A0EoMEx}\\[2ex]
\nabla\times\delta\vec{B} &=& -ie\left(\varphi_0^*\nabla\varphi_0 -\varphi_0\nabla\varphi_0^*\right) -2e(e\vec{A}_0 +\nabla\alpha_0)|\varphi_0|^2 \, , \label{deltaAEoMEx}
\eea
\end{subequations}
where we have abbreviated 
\bea \label{Ddef}
{\cal D}_0 &\equiv& \Delta - 2i(e\vec{A}_0+\nabla\alpha_0)\cdot\nabla-i\nabla\cdot(e\vec{A}_0+\nabla\alpha_0)-(e\vec{A}_0+\nabla\alpha_0)^2\non[2ex]
&&+\,(\nabla\alpha_0)^2-m_\pi^2\cos\alpha_0 \, .
\eea
The equation of motion (\ref{alpha0EoMEx}) is solved by the CSL solution, as discussed in Sec.\ \ref{sec:csl1}, while Eq.\ (\ref{A0EoMEx}) is trivially solved by the constant leading-order field $\vec{B}_0 = B_{c2}\vec{e}_z$. The other equations will now be discussed step by step, starting with Eq.\ (\ref{phi0EoMEx}). 

\subsection{Compute $\varphi_0$}
\label{sec:phi0}

With $e\vec{A}_0+\nabla\alpha_0 = eB_{c2}x\vec{e}_y$ and Eq.\ (\ref{dza0}), we write Eq.\ (\ref{phi0EoMEx}) as 
\be
0 = \left\{\partial_x^2+\partial_y^2+\partial_z^2 -2ieB_{c2}x\partial_y -e^2B_{c2}^2 x^2 +\frac{m_\pi^2}{p^2}\left[4+p^2-6p^2{\rm sn}^2(\bar{z},p^2)\right]\right\} \varphi_0 \, .
\ee
This equation is solved by a product ansatz, where the $z$-dependence is given by the wave function of the state that becomes unstable, 
\be
\varphi_0(\vec{r}) = \phi_0(x,y)f_0(\bar{z}) \, , 
\ee
with $f_0(\bar{z})$ from Eq.\ (\ref{f0}). 
With $B_{c2}$ from Eq.\ (\ref{nudef}) this yields 
\be
0 = (\partial_x^2+\partial_y^2 -2ieB_{c2}x\partial_y -e^2B_{c2}^2 x^2 +eB_{c2})\phi_0 \, . 
\ee
This is the usual differential equation obtained at the second critical field of a type-II superconductor \cite{Abrikosov:1957classic,Kleiner:1963Bulk,Evans:2022hwr}, such that we can immediately write down the solution 
\be\label{phi0}
\phi_0(x,y) = \sum_{n=-\infty}^\infty C_n e^{inqy} e^{-\frac{(x-nq\xi^2)^2}{2\xi^2}}\, , 
\ee
where $C_n$ are complex coefficients, $q$ is the wave number, and 
\be\label{xi}
\xi \equiv \frac{1}{\sqrt{eB_{c2}}}
\ee
is the coherence length. We shall only consider solutions that have discrete translational invariance and thus produce a regular 2D Abrikosov lattice, namely 
\be \label{Cn}
C_n = \left\{\begin{array}{cc} C & \mbox{$\;\;\;$ for even $n$} \\
\pm iC & \mbox{$\;\;\;$ for odd $n$} \end{array}\right. \, , 
\ee
with $C\in \mathbb{C}$ and $|C|$ to be determined. The unit cell in the $x$-$y$-plane is given by the rectangle with side lengths
\be \label{LxLy}
L_x = 2q\xi^2 \, , \qquad L_y = \frac{2\pi}{q} \, .
\ee
The lattice structure can be varied continuously by changing $q$. It is convenient to work instead with the dimensionless parameter
\be \label{adef}
a\equiv  \frac{L_x}{L_y} = \frac{q^2\xi^2}{\pi}  \, . 
\ee
Restricting ourselves to $a\in [0,1]$, we have a quadratic lattice for $a=1$ and a hexagonal lattice for $a=1/\sqrt{3}$. (The range $a\in[1,\infty]$ gives the same structures with $x$ and $y$ exchanged.) The hexagonal structure minimizes the free energy in a  standard type-II superconductor \cite{Kleiner:1963Bulk}. We shall see later that this is still the case in our system, despite the additional lattice structure in the $z$ direction.
For the following we introduce the notation for averaging any field $f(\vec{r})$ over the $x$-$y$ plane, 
\be \label{xyAv}
\langle f(\vec{r})\rangle_{x,y} = 
\frac{1}{L_x L_y}\int_0^{L_x}dx\int_0^{L_y} dy \, f(\vec{r}) \, . 
\ee
In order to solve the remaining equations of motion and to compute the free energy, it is convenient to apply a Fourier decomposition. Eventually, we will need to go to Fourier space with respect to all three spatial dimensions. Here we start by a two-dimensional Fourier series for the 2D Abrikosov lattice given by $\phi_0$. We will need the Fourier series of 
\be \label{omegadef}
\omega(x,y) \equiv |\phi_0(x,y)|^2 \, .  
\ee
We denote the Fourier transform by $\hat{\omega}(n,m) = \hat{\omega}(\vec{k}_\perp)$ with 
\be
\omega(x,y) = \sum_{\vec{k}_\perp} e^{i\vec{k}_\perp\cdot\vec{r}} \hat{\omega}(\vec{k}_\perp) \, , 
\ee
where $\vec{k}_\perp = (k_x,k_y)$ with 
\be \label{kxky}
k_x=\frac{\pi(n+2m)}{q\xi^2} \, , \qquad k_y = qn \, .
\ee
Hence, the sum over $\vec{k}_\perp$ stands for a double sum over $n,m\in [-\infty,\infty]$. The inverse transformation is 
\be\label{FourierInv}
\hat{\omega}(\vec{k}_\perp) = \langle e^{-i\vec{k}_\perp\cdot\vec{r}} \omega(x,y)\rangle_{x,y} \, ,
\ee
and we have the orthogonality relation 
\be
\langle e^{i\vec{k}_{1,\perp}\cdot\vec{r}}e^{-i\vec{k}_{2,\perp}\cdot\vec{r}}\rangle_{x,y} = \delta_{n_1n_2}\delta_{m_1m_2} \, .
\ee
In Appendix \ref{app:fourier} we compute the Fourier transform of $\omega(x,y)$ explicitly, with the result (obtained already in Ref.\ \cite{Kleiner:1963Bulk} for the hexagonal lattice)
\be \label{omeganm}
\hat{\omega}(n,m) = \frac{|C|^2}{\sqrt{a}}(\pm 1)^n(-1)^{nm}e^{-\frac{\pi}{a}\left(m^2+nm+\frac{a^2+1}{4}n^2\right)} \, .
\ee
Here, the $\pm$ originates from the two possible sign choices in Eq.\ (\ref{Cn}). This choice plays no role for any of the physics and thus we may simply continue with the upper sign, in which case we have $\hat{\omega}(\vec{k}_\perp) = \hat{\omega}(-\vec{k}_\perp)$.  
We can use this Fourier transform to compute 
\be
\label{beta}
\beta \equiv \frac{\langle |\phi_0|^4\rangle}{\langle |\phi_0|^2\rangle^2} = \sum_{n,m}\frac{\hat{\omega}^2}{\hat{\omega}_0^2} = \sum_{n,m}e^{-\frac{2\pi}{a}\left(m^2+nm+\frac{a^2+1}{4}n^2\right)} \, ,
\ee
where we have abbreviated 
\be \label{om0def}
\hat{\omega}_0\equiv \hat{\omega}(0,0) =\frac{|C|^2}{\sqrt{a}}\, .
\ee
The quantity $\beta$ was introduced in Abrikosov's original paper \cite{Abrikosov:1957classic}. It only depends on $a$ and plays an important role in determining the preferred lattice structure. With the help of the Fourier transform of $\omega$, the calculation of $\beta$ is particularly simple. For a direct calculation in position space see for instance Appendix A in Ref.\ \cite{Evans:2022hwr}, where the result is expressed in terms of the Jacobi theta function instead of the double sum of Eq.\ (\ref{beta}). We will use $\beta$ also in our final result for the free energy but will find that, in contrast to the standard Ginzburg-Landau superconductor, additional functions of $a$ occur.

In the following, we will need the full 3D Fourier transform of the charged condensate squared $|\varphi_0(\vec{r})|^2$. Having discussed the transverse directions, we now turn to the longitudinal direction and denote 
\be
s(z) \equiv f_0^2(z) \, .
\ee
We write the Fourier series in the $z$ direction as 
\be
s(z) = \sum_{k_z} e^{ik_z z} \hat{s}(k_z) \, , 
\ee
where 
\be
k_z = \frac{m_\pi\pi}{pK(p^2)}\ell \, , 
\ee
such that the sum over $k_z$ is a sum over $\ell\in [-\infty,\infty]$.
The inverse transformation is 
\be \label{sinv}
\hat{s}(\ell) = \langle  e^{-i\bar{k}_z\bar{z}} s(\bar{z})\rangle_z \, , 
\ee
where we have introduced the dimensionless wave number $\bar{k}_z = pk_z/m_\pi$, and the orthogonality relation is
\be
\langle e^{i\bar{k}_{1z}\bar{z}} e^{-i\bar{k}_{2z}\bar{z}}\rangle_z = \delta_{\ell_1\ell_2} \, .
\ee
Since $s$ is even in $z$ we can work with the somewhat simpler form 
\be
s(z) = 1 + 2\sum_{\ell=1}^\infty \cos(k_z z) \hat{s}(\ell) \, . 
\ee
Here we have used $\hat{s}(0)=1$ due to the normalization (\ref{normalize}). All other Fourier components have to be computed numerically from Eq.\ (\ref{sinv}).

\subsection{Compute $\delta B$}

The goal of this section is to calculate the correction to the magnetic field due to the charged pion condensate. Since the charged pions couple to the neutral pion field, which, in turn, is modulated in the $z$ direction already in the pure CSL phase, we can expect a modulation of the magnetic field in the $z$ direction as well. Because of the Maxwell equation $\nabla\cdot\vec{B}=0$, this inevitably creates nonzero (and non-constant) $x$ and $y$ components of the magnetic field. These expectations are borne out in the following solution to Eq.\ (\ref{deltaAEoMEx}). The components of this vector equation are  
\begin{subequations}\label{dAs0}
\bea
\partial_x\nabla\cdot\delta\vec{A}  -\Delta\delta A_x  &=& -es\partial_y\omega \, , \\[2ex]
\partial_y\nabla\cdot\delta\vec{A}-\Delta\delta A_y   &=& es\partial_x\omega \, , \\[2ex]
\partial_z\nabla\cdot\delta\vec{A}   -\Delta\delta A_z   &=& 0 \,.
\eea
\end{subequations}
We solve these coupled partial differential equations in Fourier space. Separating a term linear in $x$ to account for the correct boundary conditions of the magnetic field and combining the longitudinal and transverse Fourier series discussed above to a 3D Fourier series, we write
\be \label{delAhat}
\delta\vec{A}(\vec{r}) = cx\vec{e}_y +\sum_{\vec{k}}e^{i\vec{k}\cdot\vec{r}}\delta\hat{\vec{A}}(\vec{k}) \, , 
\ee
where $c$ is a constant, to be determined later, and where $\vec{k}=(\vec{k}_\perp,k_z)$. Employing Coulomb gauge, $\nabla\cdot\delta\vec{A}=0$, Eqs.\ (\ref{dAs0}) can now easily be solved in Fourier space to obtain
\begin{subequations}
\bea
\delta \hat{A}_x(\vec{k}) &=& -\frac{ik_y}{k^2}e\hat{s}(k_z)\hat{\omega}(\vec{k}_\perp) \, , \\[2ex]
\delta \hat{A}_y(\vec{k}) &=& \frac{ik_x}{k^2}e\hat{s}(k_z)\hat{\omega}(\vec{k}_\perp) \, , \\[2ex]
\delta \hat{A}_z(\vec{k}) &=& 0 \, .
\eea
\end{subequations}
As a consistency check, one confirms that the gauge condition is obeyed in Fourier space, $k_x\delta\hat{A}_x + k_y\delta\hat{A}_y + k_z\delta\hat{A}_z=0$. With the help of Eq.\ (\ref{delAhat}) we can write the magnetic field as
\be
\delta\vec{B}(\vec{r}) = c\vec{e}_z+\sum_{\vec{k}}e^{i\vec{k}\cdot\vec{r}}\delta\hat{\vec{B}}(\vec{k}) \, , 
\ee
where $\delta\hat{\vec{B}} = i\vec{k}\times \delta\hat{\vec{A}}$. Consequently, the magnetic field components in Fourier space are 
\begin{subequations} \label{dBCSL}
\bea
\delta \hat{B}_x(\vec{k}) &=& \frac{k_xk_z}{k^2}e\hat{s}(k_z)\hat{\omega}(\vec{k}_\perp)  \, , \\[2ex]
\delta \hat{B}_y(\vec{k})   &=& \frac{k_y k_z}{k^2}e\hat{s}(k_z)\hat{\omega}(\vec{k}_\perp) \, , \\[2ex]
  \delta \hat{B}_z(\vec{k})   &=& -\frac{k_\perp^2}{k^2}e\hat{s}(k_z)\hat{\omega}(\vec{k}_\perp) \, .
\eea
\end{subequations}
To implement the boundary condition, we use Eq.\ (\ref{BB0}) to compute $\langle B\rangle\equiv \langle B_z\rangle_{x,y} = c+B_{c2}-e\hat{\omega}_0$, i.e., 
\be \label{cdef}
c=\langle B\rangle-B_{c2}+e\hat{\omega}_0 \, . 
\ee
Inserting this expression for $c$, we can write the magnetic field components in position space as
\begin{subequations}\label{Bxyz}
\bea
B_x(\vec{r}) &=&   2ie\sum_{\vec{k}_\perp}\sum_{\ell=1}^\infty e^{i\vec{k}_\perp\cdot\vec{r}}\sin(k_z z)\frac{k_x k_z}{k^2}\hat{s}\hat{\omega} \, , \\[2ex]
B_y(\vec{r}) &=&   2ie\sum_{\vec{k}_\perp}\sum_{\ell=1}^\infty e^{i\vec{k}_\perp\cdot\vec{r}}\sin(k_z z)\frac{k_y k_z}{k^2}\hat{s}\hat{\omega} \, , \\[2ex]
B_z (\vec{r})&=& \langle B\rangle +e\hat{\omega}_0\Big(1-\sum_{\vec{k}_\perp}e^{i\vec{k}_\perp\cdot\vec{r}}\frac{\hat{\omega}}{\hat{\omega}_0}\Big) - 2e\sum_{\vec{k}_\perp}\sum_{\ell=1}^\infty e^{i\vec{k}_\perp\cdot\vec{r}} \cos(k_z z)\frac{k_\perp^2}{k^2}\hat{s}\hat{\omega}  \, .
\eea
\end{subequations}
\begin{figure} [t]
\begin{center}
\hbox{\includegraphics[width=0.5\textwidth]{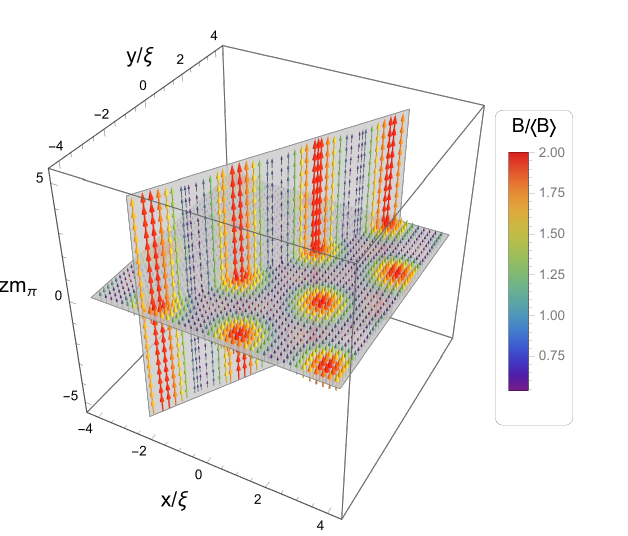}
\includegraphics[width=0.5\textwidth]{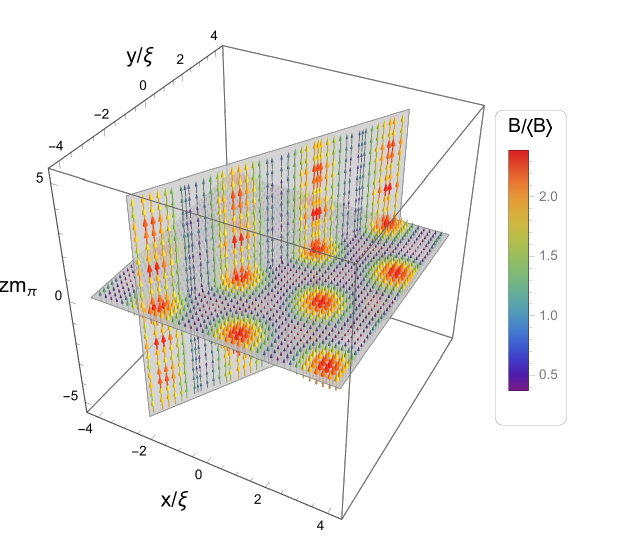}}
\hbox{\includegraphics[width=0.5\textwidth]{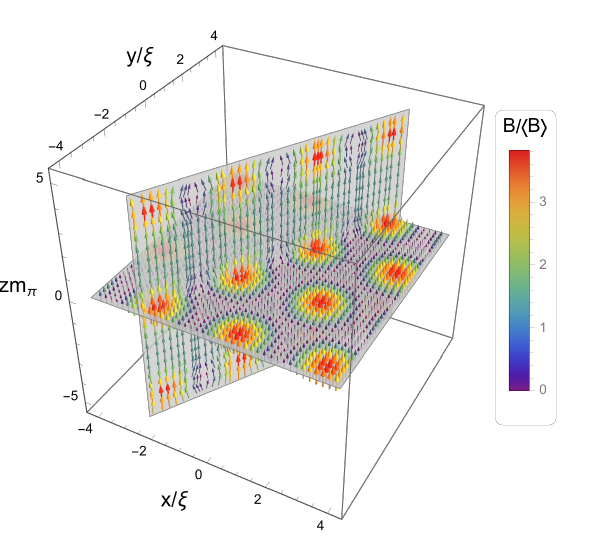}
\includegraphics[width=0.5\textwidth]{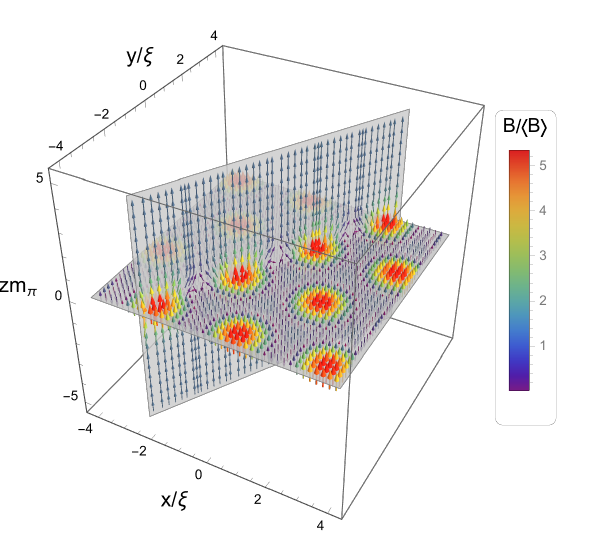}}
\caption{Dimensionless magnetic field $\vec{B}/\langle B\rangle$ from Eq.\ (\ref{Bxyz}) for the hexagonal lattice $a=1/\sqrt{3}$.  From top left to bottom right, the elliptic modulus is $p=0.1,0.5,0.9,0.99$, which gives a sequence from (near) the chiral limit to (near) the limit of a single domain wall. For all plots, we have set $\langle B\rangle/(e\hat{\omega}_0) =1$, which exaggerates the effect of deviations from a constant magnetic field for illustrative purposes. Our expansion assumes these deviations to be small.}
\label{fig:B}
\end{center}
\end{figure} 
In contrast to the lowest-order contribution to the charged pion condensate $\varphi_0(\vec{r})$, the magnetic field correction $\delta\vec{B}(\vec{r})$ does not factorize into fields depending on longitudinal and transverse directions separately. With the help of Eq.\ (\ref{Bxyz}) we can evaluate $\vec{B}/\langle B\rangle$ numerically after choosing values for $a$, $\langle B\rangle/(e\hat{\omega}_0)$, and $p$. In Fig.\ \ref{fig:B} we plot the result for the hexagonal lattice $a=1/\sqrt{3}$ and $\langle B\rangle/(e\hat{\omega}_0)=1$. Since $\vec{B}/(e\hat{\omega}_0)$ is of order $1/\epsilon^2$, the chosen value is somewhat small, and our approximation is strictly speaking questionable for this value.  The benefit of this choice is that the deviations of $\vec{B}$ from the constant field $\vec{B}_0$ become more visible. The final parameter, the elliptic modulus $p$, is used to show different scenarios in the range between the chiral limit and the domain wall. To this end, we have chosen four different values of $p$. In the upper left panel we show a case close to the chiral limit, which reproduces the result of Ref.\ \cite{Evans:2022hwr}: The magnetic field is confined in flux tubes and (nearly) constant in the $z$ direction, as in an ordinary type-II superconductor. The distance between the flux tubes is set by the scale $\xi$. As we increase the elliptic modulus we see that the $z$-dependence becomes non-trivial, with the magnetic field enhanced in bubble-like regions. The lower right panel is close to the domain wall configuration, where the layers in the $z$ direction are well separated and the magnetic field between them is nearly constant.

\subsection{Compute $\delta\alpha$}

Next, we turn to Eq.\ (\ref{deltaalphaEoMEx})  to compute $\delta\alpha$. We first separate a topological (= nonzero winding) contribution from the contribution that we will write as a Fourier series. For fixed $\mu$, the CSL solution $\alpha_0$ with elliptic modulus $p$ is valid at $B_{c2}$. We are interested in the new $\alpha_0+\delta\alpha$  at $\langle B\rangle$. Therefore, we introduce the elliptic modulus 
$p+\delta p$, 
which reproduces the CSL solution at $\langle B\rangle$.  
Using Eq.\ (\ref{minik}) for $p$ at $B_{c2}$ and for $p+\delta p$ at $\langle B\rangle$, we find 
\be \label{delp}
\delta p = -\frac{pE(p^2)}{K(p^2)}\frac{\langle B\rangle - B_{c2}}{B_{c2}} +{\cal O}(\epsilon^4)\, .
\ee
The topological contribution to $\delta \alpha$ can now be obtained by expanding $\alpha_0$ at elliptic modulus $p+\delta p$ for small $\delta p$. This gives the ${\cal O}(1)$ contribution $\alpha_0$ at elliptic modulus $p$ plus an ${\cal O}(\epsilon^2)$ contribution that we denote by $\alpha_1\delta p$. Together with a non-topological contribution, the total ${\cal O}(\epsilon^2)$ correction is thus 
\be \label{dalpha0}
\delta\alpha = \alpha_1\delta p +\frac{\hat{\omega}_0}{f_\pi^2}\delta \alpha_1 \, , \qquad \delta\alpha_1\equiv \sum_{\vec{k}} e^{i\vec{k}\cdot\vec{r}} \delta \hat{\alpha}(\vec{k}) \, , 
\ee
where the dimensionless factor $\hat{\omega}_0/f_\pi^2$ of order $\epsilon^2$ has been introduced for convenience, such that $\delta \hat{\alpha}$ is of order 1, and 
where
\be\label{dal1def}
\alpha_1 = \frac{\partial \alpha_0}{\partial p} = -\frac{ {\cal E}(\bar{z},p^2)\partial_{\bar{z}}\alpha_0+\partial^2_{\bar{z}}\alpha_0}{p(1-p^2)} \, ,
\ee
 with the Jacobi epsilon function ${\cal E}$. (The derivative with respect to $p$ is taken at fixed $z$, not fixed $\bar{z}$.)  With 
\be \label{deps}
\partial_{\bar{z}}{\cal E} = \frac{(\partial_{\bar{z}}\alpha_0)^2}{4} 
\ee
we easily verify that $\alpha_1$ fulfills the differential equation
\be\label{ODEal1}
(\partial^2_{\bar{z}}-p^2\cos\alpha_0)\alpha_1 = 0 \, .
\ee
We are now prepared to solve Eq.\ (\ref{deltaalphaEoMEx}). To this end, we first compute 
\be
(\Delta\alpha_0+2\nabla\alpha_0\cdot\nabla) |\varphi_0|^2   =  m_\pi^2 \omega(x,y)t(\bar{z})\, , 
\ee
where 
\bea \label{tdef}
&&t(\bar{z}) \equiv \frac{1}{p^2} (\partial_{\bar{z}}^2\alpha_0 +2\partial_{\bar{z}}\alpha_0\partial_{\bar{z}})s(\bar{z}) \non[2ex]
&&=\frac{9\sin\alpha_0}{N(p)}\left[\frac{\sqrt{p^4-p^2+1}+1-2p^2}{3p^2}+\sin^2\frac{\alpha_0}{2}\right]\left[\frac{\sqrt{p^4-p^2+1}+25-26p^2}{27p^2}+\sin^2\frac{\alpha_0}{2}\right] \, . \non
&&
\eea
This function is antisymmetric, $t(\bar{z}) = -t(-\bar{z})$, and thus its Fourier components can be computed numerically via 
\be
\hat{t}(\ell) = -\frac{i}{K(p^2)}\int_{0}^{K(p^2)} d\bar{z}\, \sin(\bar{k}_z\bar{z}) t(\bar{z}) \, . 
\ee
In particular, the zero mode vanishes, $\hat{t}(0)=0$. Furthermore, we abbreviate the function 
\be
u(\bar{z}) \equiv \cos\alpha_0 \, , 
\ee
whose Fourier components $\hat{u}(\ell)$ we compute numerically.

\begin{figure} [t]
\begin{center}
\includegraphics[width=0.5\textwidth]{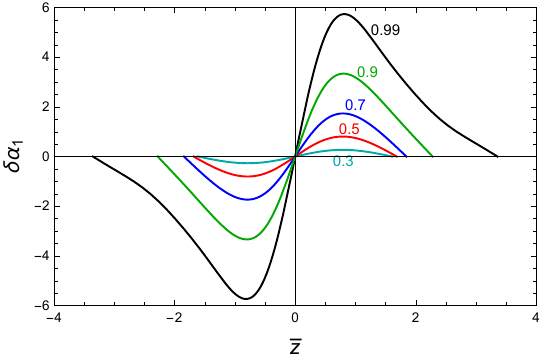}
\caption{Non-topological contribution to $\delta\alpha$ as a function of $\bar{z}$ for different elliptic moduli $p$, as indicated by the labels, and $a=1/\sqrt{3}$ at the point $(x,y) = \pi^{1/2}/2\,(3^{-1/4},3^{1/4})$, corresponding to a local extremum of the 2D hexagonal lattice. In each case the curve is plotted over one period $-K(p^2)<\bar{z}<K(p^2)$.}
\label{fig:dalpha}
\end{center}
\end{figure}

Inserting all this into Eq.\ (\ref{deltaalphaEoMEx}), we obtain in Fourier space,
\bea\label{eqdal}
0 &=& \frac{k^2}{m_\pi^2}\delta\hat{\alpha}(n,m,\ell)+\sum_{\ell'=1}^\infty [\hat{u}(\ell-\ell')-\hat{u}(\ell+\ell')]\delta\hat{\alpha}(n,m,\ell')+\frac{\hat{\omega}(n,m)}{\hat{\omega}_0}\hat{t}(\ell)  \, .
\eea 
Importantly, the topological contribution has dropped out due to Eq.\ (\ref{ODEal1}). We have assumed the non-topological part to be odd in $\bar{z}$, i.e., $\delta\hat{\alpha}(\ell) = -\delta\hat{\alpha}(-\ell)$, which gives a consistent solution to the differential equation. For the practical calculation, it is useful to note that  
\bea
\frac{k^2}{m_\pi^2} &=&  \frac{1}{p^2}\left[\frac{4\pi}{a}\left(m^2+nm+\frac{1+a^2}{4}n^2\right)\left(2-p^2+2\sqrt{p^4-p^2+1}\right)+\frac{\pi^2\ell^2}{K^2(p^2)}\right] 
\eea
only depends on the geometric parameter $a$ and the elliptic modulus $p$ (besides the integers $n,m,\ell$), and not explicitly on $m_\pi$. Therefore, for given $a$ and $p$, Eq.\ (\ref{eqdal}) yields the Fourier modes $\delta\hat{\alpha}(n,m,\ell)$ and thus the non-topological contribution $\delta\alpha_1$. Due to the term with the convolution, for each transverse mode $(n,m)$ we have to solve a coupled set of linear equations for the longitudinal Fourier modes $\ell$. In the practical calculation, we must, of course, restrict ourselves  to a truncation, say $n,m\in[-n_\perp,n_\perp]$ and $\ell\in [-n_z,n_z]$.  It turns out that convergence to the final result is already achieved to a very good accuracy for $n_\perp$ of about 3 or 4 for all quantities we consider, while it is usually sufficient to choose $n_z$ of the order of 10 (except for elliptic moduli $p$ very close to 1, where we need up to $n_z\sim 100$). We have also verified our numerical results by inserting $\delta\alpha_1(\vec{r})$, obtained from its Fourier components, back into the differential equation (\ref{deltaalphaEoMEx}). In Fig.\ \ref{fig:dalpha} we show $\delta\alpha_1$ for $a=1/\sqrt{3}$ and different values of $p$ at a fixed point $(x,y)$ as a function of $\bar{z}$. This plot is shown mostly for replicability of our calculation, $\delta\alpha_1$ is one of the ingredients needed later for our physical results. 

We have now solved 5 of the 6 equations of motion (\ref{EoMEx}). The equation for $\delta \varphi$ (\ref{deltaphiEoMEx}) does not need to be solved explicitly for our purposes. We will make use of it implicitly in the calculation of the free energy in Sec.\ \ref{sec:free}.

\subsection{Crystalline baryon density} 
\label{sec:baryon}

With the results of the previous subsections we can now calculate the local baryon number density $n_B$ (\ref{nblocal}). We shall evaluate $n_B$ consistently up to order $\epsilon^2$. As we see from Eq.\ (\ref{nblocal}), $n_B$ has a magnetic and a vorticity contribution (this structure from the chiral anomaly reflects the well-known chiral magnetic and chiral vortical effects \cite{Landsteiner:2016led}).  The magnetic contribution contains the leading-order CSL part plus $\epsilon^2$ corrections, 
\bea \label{nBmag}
\frac{\nabla\alpha\cdot e\vec{B}}{4\pi^2}  &\simeq& 
\frac{\partial_z\alpha_0 eB_{c2}}{4\pi^2} +\left[\partial_z\alpha_0-\frac{pE(p^2)}{K(p^2)}\partial_z\alpha_1\right]\frac{e(\langle B\rangle -B_{c2})}{4\pi^2}+\frac{e\partial_z\alpha_0}{4\pi^2}\Bigg(e\hat{\omega}_0 + \sum_{\vec{k}}e^{i\vec{k}\cdot\vec{r}} \delta\hat{B}_z\Bigg) \non[2ex]
&&+\,\frac{e\hat{\omega}_0 B_{c2}}{4\pi^2 f_\pi^2}\sum_{\vec{k}}ik_ze^{i\vec{k}\cdot\vec{r}}\delta\hat{\alpha}\, , 
\eea
where the Fourier transforms $\delta\hat{B}_z$ and $\delta\hat{\alpha}$ are obtained from Eqs.\ (\ref{dBCSL}) and (\ref{eqdal}).

The vorticity contribution is of order $\epsilon^2$, and 
we first note that $\varphi$, $\alpha$, and $\vec{A}$ in the three-current $\vec{j}$ can all be replaced by their lowest-order contributions. With the Maxwell law (\ref{deltaAEoMEx}) we can therefore use $\vec{j} = \nabla\times \delta\vec{B}$ and thus $\nabla\times \vec{j}= -\Delta\delta\vec{B}$ in the calculation of the baryon density, which yields  
\bea
\frac{\nabla\alpha\cdot(\nabla\times\vec{j})}{4\pi^2ef_\pi^2} \simeq \frac{\partial_z\alpha_0}{4\pi^2 ef_\pi^2}\sum_{\vec{k}}k^2e^{i\vec{k}\cdot\vec{r}}\delta\hat{B}_z \, .
\eea
For a quantitative plot of the baryon density it is necessary to anticipate a result of Sec.\ \ref{sec:eomega0}, namely Eq.\ (\ref{EOM0}), which allows us to evaluate $n_B$ for given values of $a$, $p$, $m_\pi$, and the  smallness parameter $\epsilon$ (\ref{epsilon}). 

\begin{figure} [t]
\begin{center}
\hbox{\includegraphics[width=0.5\textwidth]{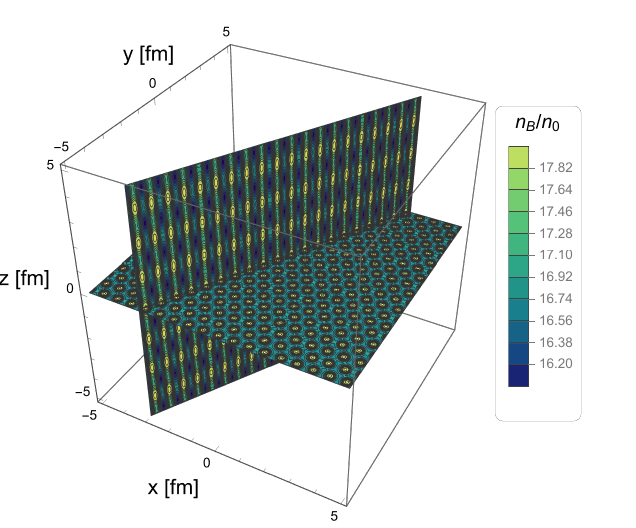}\includegraphics[width=0.5\textwidth]{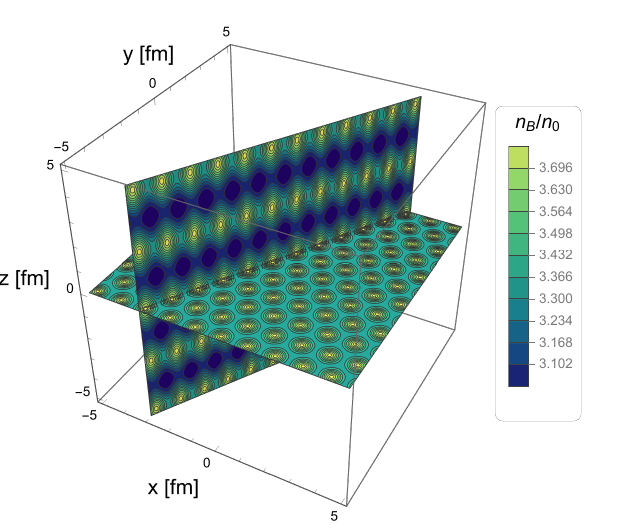}}
\hbox{\includegraphics[width=0.5\textwidth]{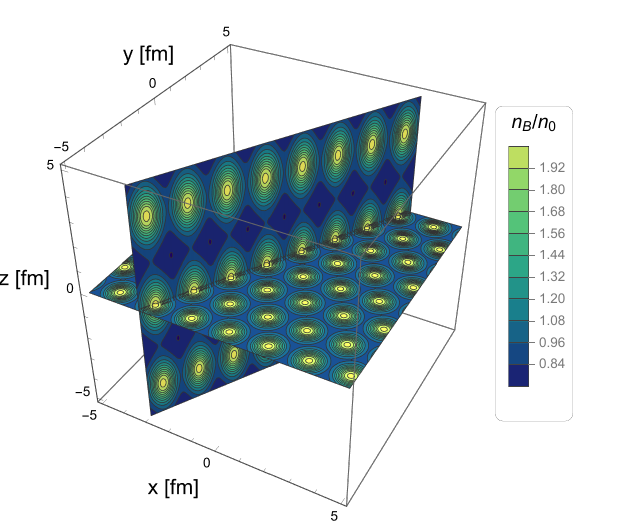}\includegraphics[width=0.5\textwidth]{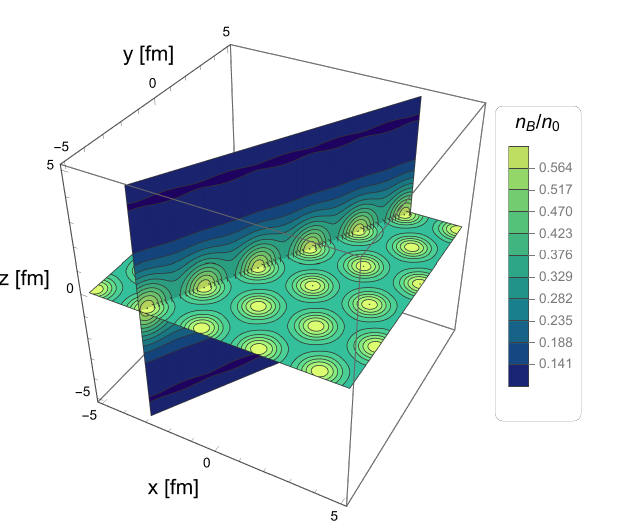}}
\caption{Baryon density in units of nuclear saturation density $n_0$ for the physical pion mass $m_\pi=140\, {\rm MeV}$, hexagonal lattice in the transverse plane, $a=1/\sqrt{3}$, and elliptic moduli $p=0.3,0.5,0.7,0.999$, from top left to bottom right, which correspond to four points on the CSL instability line $(\mu\,  [{\rm MeV}],eB_{c2}\, [{\rm GeV}^2]) \simeq (366,0.83), (628,0.28),(889,0.13),(1016,0.059)$. The lattices are computed away from the instability line with $\epsilon = 0.1$ (first three plots) and $\epsilon=0.3$ (bottom right plot). Here and in all following numerical results we have used the value of the pion decay constant $f_\pi=92\, {\rm MeV}$. The vertical direction is distinguished by the magnetic field, which points (predominantly) into the $z$ direction, see Fig.\ \ref{fig:B}. The variation of $n_B$ in the transverse direction is due to charged pions, whose vorticity contributes to the topological baryon number, in addition to a magnetic field contribution.}
\label{fig:nB}
\end{center}
\end{figure}

The results are shown in Fig.\ \ref{fig:nB} for the physical pion mass and a hexagonal lattice.  We show the baryon density in units of saturation density of symmetric nuclear matter, $n_0=0.15\, {\rm fm}^{-3}$, and use physical units for the spatial directions, in particular employing  Eq.\ (\ref{xi}) for the coherence length. The average value of the baryon number for each plot is determined by the CSL phase at the specific point $(\mu, eB_{c2})$ on the instability curve, see Eq.\ (\ref{nBCSL}). Also, the CSL phase already gives rise to variations in the baryon number in the $z$ direction. The variations in the transverse direction, however, only arise due to the presence of the charged pions as we move away from the CSL instability curve by an amount given by $\epsilon$. We have set $\epsilon=0.1$ for all plots except for the bottom right one, where a larger value, $\epsilon=0.3$, is needed to make the lattice structure clearly visible. In the upper left plot we see that for very large average magnetic field and thus very large average baryon density, baryon number is localized in tube-like structures, elongated in the direction of the magnetic field. (For $eB_{c2}\to\infty$ the result approaches the chiral limit, where the baryons are localized in tubes without any variation in $z$.) As we move to smaller magnetic fields, the lattice becomes coarser and the size of the baryon lumps becomes larger, approaching the size of actual baryons as we approach densities $n_B\sim n_0$. The bottom right panel shows a sheet-like structure, where baryon number is confined in almost spherical bubbles within a domain wall that is well separated in the $z$ direction from the next sheet. As the 
results of Sec.\ \ref{sec:df} will show, only the first three panels of Fig.\ \ref{fig:nB} depict configurations that are stable in our perturbative approach.

\section{Free energy of the 3D pion crystal}
\label{sec:free}

\subsection{General expression}
\label{sec:general}

After having constructed the 3D pion crystal, we need to compute its free energy to determine whether it provides a solution to the CSL instability. Before inserting our solution into the effective potential, we can bring the free energy into a convenient form in exactly the same way as for the ordinary Abrikosov lattice and for the chiral limit of Ref.\ \cite{Evans:2022hwr}. Starting from   (\ref{Omega0}), we derive \cite{Evans:2022hwr}
\bea \label{FreeEnergyEx}
F \simeq \frac{1}{V}\int d^3\vec{r}\left\{\frac{B^2}{2}+\frac{f_\pi^2}{2}[(\nabla\alpha)^2-2m_\pi^2(\cos\alpha-1)]-\lambda_*|\varphi_0|^4-\frac{e\mu}{4\pi^2}\nabla\alpha\cdot\vec{B}
\right\} \, ,
\eea
where $V$ is the volume of the system, where we have subtracted the vacuum contribution, as in Eq.\ (\ref{OmegaCSL}), and where we have introduced the effective coupling 
\be
\lambda_*\equiv \frac{\langle 
(\nabla|\varphi_0|^2)^2\rangle +m_\pi^2\langle|\varphi_0|^4 \cos\alpha_0 \rangle}{2f_\pi^2\langle|\varphi_0|^4\rangle} \, .
\ee
Here, $\langle - \rangle$ denotes averaging over all three spatial directions, combining the integrals (\ref{zAv}) and (\ref{xyAv}). 
In the textbook treatment of a Ginzburg-Landau superconductor, this coupling corresponds to a constant in front of the quartic term of the complex scalar field. As written, the free energy (\ref{FreeEnergyEx}) contains terms of all orders in $\epsilon$ through $\alpha$ and $B$, while the term $\lambda_*|\varphi_0|^4$ is of order $\epsilon^4$. This expression will turn out to be useful because we are eventually interested in the ${\cal O}(\epsilon^4)$ free energy difference between our crystal and the CSL phase.  

The final ingredient for the calculation is the orthogonality relation 
\be
0 = \int d^3\vec{r}\, \varphi_0^*{\cal D}_0\delta\varphi \, ,
\ee
which, again following the steps of Ref.\ \cite{Evans:2022hwr}, can be recast in the form 
\bea 
0&=& \int d^3\vec{x}\left[2\lambda_*|\varphi_0|^4-(\delta\vec{A}+\nabla\delta\alpha/e)\cdot(\nabla\times\delta\vec{B}) -|\varphi_0|^2(\Delta\alpha_0+2\nabla\alpha_0\cdot\nabla)\delta\alpha\right] \, , \label{intAdB}
\eea
where Eqs.\ (\ref{deltaphiEoMEx}) and (\ref{deltaAEoMEx}) have been used. In the following subsections we will compute $\lambda_*$, evaluate Eq.\ (\ref{intAdB}) to extract an expression for $e\hat{\omega}_0$ (which corresponds to determining the coefficient of the Abrikosov lattice $|C|$), and then use these results to compute the free energy $F$ (\ref{FreeEnergyEx}). 

\subsection{Compute  $\lambda_*$}
\label{sec:lambda}

The calculation of $\lambda_*$ can be done analytically. Each of the relevant functions factorizes into longitudinal and transverse parts, \begin{subequations}\allowdisplaybreaks
\bea
\langle (\nabla|\varphi_0|^2)^2\rangle &=&\langle s^2 \rangle_z \langle (\nabla\omega)^2\rangle_{x,y} + 
\langle (\partial_z s)^2 \rangle_z \langle \omega^2\rangle_{x,y} \\[2ex]
\langle|\varphi_0|^4 \cos\alpha_0 \rangle &=& \langle s^2 \cos\alpha_0 \rangle_z \langle \omega^2\rangle_{x,y} \\[2ex]
\langle|\varphi_0|^4  \rangle &=& \langle s^2 \rangle_z \langle \omega^2\rangle_{x,y} \, .
\eea
\end{subequations}
Moreover, we use the result 
\be
\frac{\langle(\nabla\omega)^2\rangle_{x,y}}{\langle\omega^2\rangle_{x,y}}=\frac{1}{\xi^2} \, ,
\ee
whose derivation can be found in Appendix C of Ref.\ \cite{Evans:2022hwr}. 
Then, we find
\be\label{lamstar}
\lambda_* = \frac{1}{2f_\pi^2} \left[eB_{c2}+\frac{\langle (\partial_z s)^2 \rangle_z +m_\pi^2\langle s^2 \cos\alpha_0 \rangle_z}{\langle s^2 \rangle_z} \right] = \frac{eB_{c2}}{2f_\pi^2} {\cal Q}(p)\, ,
\ee
where we have introduced the function
\be \label{Qdef}
{\cal Q}(p) \equiv 1+\frac{1}{b p^2}\frac{(q_1+q_2\sqrt{p^4-p^2+1})E(p^2)+(q_3+q_4\sqrt{p^4-p^2+1})K(p^2)}{(q_5+q_6\sqrt{p^4-p^2+1})E(p^2)+(q_7+q_8\sqrt{p^4-p^2+1})K(p^2)} \, , 
\ee
with the polynomials
\begin{subequations}\allowdisplaybreaks
\bea
q_1(p) &\equiv& 3(12-24p^2+61p^4-49p^6+22p^8) \, , \\[2ex]
q_2(p) &\equiv&6(6-9p^2+89p^4-43p^6) \, , \\[2ex]
q_3(p) &\equiv& -36+90p^2+2p^4-93p^6+177p^8-70p^{10} \, , \\[2ex] 
q_4(p)&\equiv&-36+72p^2-340p^4+304p^6-70p^8 \, , \\[2ex]
q_5(p)&\equiv& -15(2-3p^2-3p^4+2p^6) \, ,\\[2ex]
q_6(p)&\equiv&294(1-p^2+p^4) \, , \\[2ex]
q_7(p)&\equiv& 5(20-40p^2+45p^4-25p^6+14p^8) \, , \\[2ex]
q_8(p)&\equiv&14(-16+24p^2-18p^4+5p^6) \, .
\eea
\end{subequations}
The function ${\cal Q}$ interpolates between the domain wall, ${\cal Q}(1)=4/3$ and thus $\lambda_* = 2m_\pi^2/f_\pi^2$, and the chiral limit, ${\cal Q}(0)= 1$ and thus $\lambda_*= eB_{c2}/(2f_\pi^2)$.
As for the textbook Ginzburg-Landau theory, we may define the Ginzburg-Landau parameter
\be\label{kappa}
\kappa^2 = \frac{\lambda_*}{e^2} \, . 
\ee
For our system the transition from type-I to type-II behavior is not simply given by the textbook value $\kappa^2=1/2$. Nevertheless, introducing $\kappa$ is useful for compactness and to compare our result to the standard Ginzburg-Landau result.

\subsection{Compute $e\hat{\omega}_0$} 
\label{sec:eomega0}

The evaluation of Eq.\ (\ref{intAdB}) will give us an expression for $e\hat{\omega}_0$, from which  the yet unknown coefficient $|C|^2$ can be read off. We consider the various terms in Eq.\ (\ref{intAdB}) separately. First, we use that 
in Coulomb gauge $\nabla\times \delta\vec{B} = -\Delta\delta\vec{A}$ and thus, with Eq.\ (\ref{dAs0}),
\be
\nabla\delta\alpha\cdot(\nabla\times\delta\vec{B}) = es(\partial_x\omega\partial_y\delta\alpha - \partial_y\omega\partial_x\delta\alpha) \, .
\ee
With the help of our Fourier decomposition it is then easy to see that $\langle \nabla\delta\alpha\cdot(\nabla\times\delta\vec{B}) \rangle = 0$. Similarly,  
\bea
\langle \delta\vec{A}\cdot(\nabla\times\delta\vec{B})\rangle  &=& e\langle s\omega(\partial_y\delta A_x-\partial_x\delta A_y)\rangle \non[2ex]
&=& e^2\sum_{\vec{k}}\frac{k_\perp^2}{k^2} \hat{s}^2(k_z)\hat{\omega}^2(\vec{k}_\perp)-ec\hat{\omega}_0 \, ,
 \eea
 where, in the first step, we have dropped the boundary contributions, and in the second step we have used the Fourier decomposition. For the next term we recall that $\delta\alpha$ contains two contributions, see Eq.\ (\ref{dalpha0}), and we compute
\bea
\langle |\varphi_0|^2(\Delta \alpha_0 + 2\nabla\alpha_0\cdot\nabla) \delta\alpha\rangle &=& e\hat{\omega}_0[{\cal G}(p)+1](\langle B\rangle-B_{c2})\non[2ex]
&&+ \frac{m_\pi^2\hat{\omega}_0}{f_\pi^2}\sum_{\vec{k}}\hat{\omega}(\vec{k}_\perp) \hat{t}(k_z) \delta\hat{\alpha}(\vec{k})  \,. \;\;\;
\eea
The second term comes from the non-topological contribution $\delta\alpha_1$, which vanishes at the boundaries $\bar{z}=\pm K(p^2)$, and we have used partial integration before inserting the Fourier decomposition to recover the function $t(\bar{z})$ from Eq.\ (\ref{tdef}). The first term originates from the topological contribution $\alpha_1\delta p$, and one can perform all $\bar{z}$ integrals explicitly with the help of Eqs.\ (\ref{dal1def}) and (\ref{deps}). The result is expressed in terms of the function 
\bea
{\cal G}(p)&\equiv& 
\frac{2E(p^2)}{9bp^6K^2(p^2)}\frac{3r_1E(p^2)+2r_2K(p^2)}{N(p)}-1\, , 
\eea
with $N(p)$ from Eq.\ (\ref{Nnorm}), and  
\begin{subequations}
\bea
r_1(p)&\equiv& 2p^4-3p^2+4-(p^2-4)\sqrt{p^4-p^2+1} \, , \\[2ex]
r_2(p)&\equiv& p^4+2p^2-2+(p^2-2)\sqrt{p^4-p^2+1} \, .
\eea
\end{subequations}
This function has the limits ${\cal G}(0)=1$ (chiral limit) and ${\cal G}(1)=-1$ (domain wall). 
Finally, with Eqs.\ (\ref{lamstar}) and (\ref{kappa}) we 
immediately obtain 
\be
2\lambda_*\langle|\varphi_0|^4\rangle = 2e^2\kappa^2\sum_{\vec{k}}\hat{\omega}^2(\vec{k}_\perp)\hat{s}^2(k_z) \, .
\ee
We now insert everything into Eq.\ (\ref{intAdB}), use the expression for $c$ in Eq.\ (\ref{cdef}), solve the resulting equation for $e\hat{\omega}_0$, separate the $\ell=0$ mode in the Fourier series to bring the result into a convenient form, and use the definition of $\beta$ (\ref{beta}). This yields 
\be \label{EOM0}
e\hat{\omega}_0 = {\cal G}\frac{\langle B\rangle-B_{c2}}{\beta(2\kappa^2-1)+1+2{\cal H}_1  - 2{\cal H}_2} \, ,
\ee
where we have introduced   
\begin{subequations}
\bea
{\cal H}_1&\equiv& \sum_{\vec{k}_\perp}\sum_{\ell=1}^\infty\left(2\kappa^2-\frac{k_\perp^2}{k^2}\right)\frac{\hat{\omega}^2}{\hat{\omega}_0^2}\hat{s}^2 \, , \\[2ex]
{\cal H}_2 &\equiv& \frac{m_\pi^2}{e^2f_\pi^2} \sum_{\vec{k}_\perp}\sum_{\ell=1}^\infty \frac{\hat{\omega}}{\hat{\omega}_0}\hat{t}\,\delta\hat{\alpha} \, . 
\eea
\end{subequations}
These two Fourier sums need to be computed numerically as functions of $a$, $b$, $m_\pi$. Both of them vanish in the chiral limit. While this is obvious for ${\cal H}_2$ due to the prefactor $m_\pi^2$, for ${\cal H}_1$ we note that in the chiral limit, where $s(z)=1$, we have $\hat{s}(\ell)=\delta_{\ell 0}$ and thus all terms in the sum are zero.

The result (\ref{EOM0}) is important for the interpretation of our pion crystal. First of all, we recall $\hat{\omega}_0 = |C|^2/\sqrt{a}$ (\ref{om0def}). The right-hand side is a function of $a$ (characterizing the transverse lattice structure), $b$ (dimensionless version of $eB_{c2}$), and the pion mass $m_\pi$. Moreover, the right-hand side is proportional to the difference $\langle B\rangle-B_{c2}$, which is by definition of order $\epsilon^2$, while all other quantities on the right-hand side are of order 1. Hence, Eq.\ (\ref{EOM0}) gives an expression for $|C|^2$ and we confirm as a consistency check that $|C|^2$ (and $\hat{\omega}_0$) is of order $\epsilon^2$. As another check we observe that in the chiral limit, with ${\cal G}$ going to 1 and both ${\cal H}_1$ and ${\cal H}_2$ going to zero, we exactly reproduce the result of Ref.\ \cite{Evans:2022hwr}, which has the same form as for a standard superconductor. 

Since the left-hand side is greater than zero, our configuration is only a physical solution if the right-hand side is greater than zero as well. One easily checks numerically that ${\cal G}$ is positive if and only if $p<0.91792$. This is close to the point where the CSL instability curve turns around (reaches its maximal $\mu$, see Fig.\ \ref{fig:pd}). With the help of Eq.\ (\ref{munu}) one finds that this turning point is at $p\simeq 0.92751$, with $p$ being smaller than this value towards larger $eB_{c2}$. The discrepancy between these two numerical values is  due to our expansion and we will ignore the small region between them\footnote{In terms of magnetic field and chemical potential, this region is given by $3.51<eB_{c2}/m_\pi^2<3.58$ and $1061.08\, {\rm MeV}<\mu<1061.57\, {\rm MeV}$, where, for the latter, we have used the physical values of $f_\pi$ and $m_\pi$.}. Hence, the function ${\cal G}$ takes care of the two-valuedness of the CSL instability curve. As a consequence, along the entire instability curve, our solution is valid outside the stable CSL region ($\langle B\rangle >B_{c2}$ on the upper branch and $\langle B\rangle <B_{c2}$ on the lower branch) if and only if the denominator in Eq.\ (\ref{EOM0}) is positive. This denominator is not only a function of $p$, but also depends on $a$ and $m_\pi$. We shall encounter the same denominator in the calculation of the free energy and discuss its behavior at the end of the following subsection.

\subsection{When the 3D pion crystal is preferred}
\label{sec:df}

We are now prepared to compute the free energy (\ref{FreeEnergyEx}). Again, let us discuss the various terms separately.  
The pure magnetic field contribution is first rewritten with the help of Eq.\ (\ref{BB0}) as 
\bea \label{Bsquared}
\langle B^2\rangle &=& \langle \vec{B}\rangle \cdot\langle \vec{B}\rangle +\langle \delta B^2\rangle - \langle \delta\vec{B}\rangle \cdot\langle \delta\vec{B}\rangle \non[2ex]
&=& \langle B\rangle^2 +\langle \delta B_x^2\rangle+\langle \delta B_y^2\rangle+\langle \delta B_z^2\rangle-\langle \delta B_z\rangle^2 \,, 
\eea
where we have used $\langle \delta B_x\rangle_{x,y}=\langle \delta B_y\rangle_{x,y} = 0$, $\langle\vec{B}\rangle = \langle B\rangle \vec{e}_z$. 
Apart from the $\langle B\rangle^2$ contribution, all terms  are of order $\epsilon^4$. This result would not further change if we allowed for higher-order corrections to $\vec{B}$. With Eqs.\ (\ref{dBCSL}) we find
\be
\langle B^2\rangle = \langle B\rangle^2 +e^2\hat{\omega}_0^2(\beta-1)+
2e^2\sum_{\vec{k}_\perp}\sum_{\ell=1}^\infty \frac{k_\perp^2}{k^2} \hat{\omega}^2(\vec{k}_\perp)\hat{s}^2(\ell) \, .
\ee
In the chiral limit, the last term vanishes because of $\hat{s}(\ell)=\delta_{\ell 0}$.

\begin{figure} [t]
\begin{center}
\hbox{\includegraphics[width=0.5\textwidth]{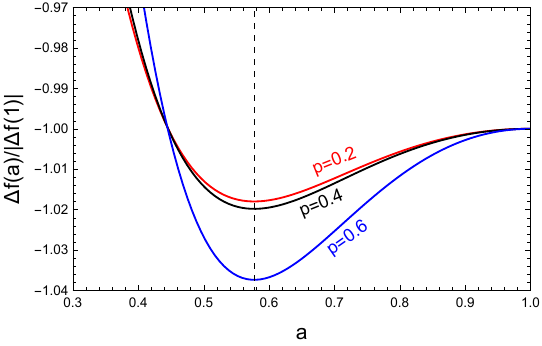}
\includegraphics[width=0.5\textwidth]{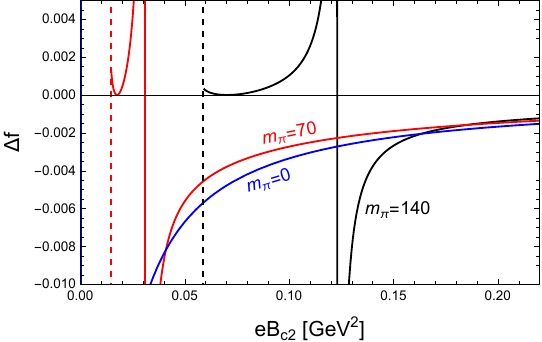}}
\caption{{\it Left panel:} Free energy difference (\ref{df}) at the physical pion mass as a function of the geometric parameter $a$, normalized to the value at $a=1$, for three different values of the elliptic modulus $p$. The vertical dashed line indicates $a=1/\sqrt{3}$, where all curves have their minimum. {\it Right panel:} Free energy difference for the hexagonal lattice $a=1/\sqrt{3}$ along the critical magnetic field, for three values of the pion mass. The dashed vertical lines indicate the smallest critical field of the CSL instability, where CSL approaches a single domain wall ($p=1$). The solid vertical lines indicate the type-I/type-II transition point, where $\Delta f$ changes sign; for critical fields larger than this point, the  pion crystal is preferred over CSL.}
\label{fig:df}
\end{center}
\end{figure}

The terms including $\alpha$ require a separate discussion. So far, we expanded $\alpha= \alpha_0+\delta\alpha$ and included a topological correction of order $\epsilon^2$ in $\delta\alpha$, together with the non-topological contribution $\delta\alpha_1$  that was computed via its Fourier modes. Since we need the free energy up to order $\epsilon^4$, this is not sufficient. One can easily show that any non-topological $\epsilon^4$ contribution that vanishes at the boundaries of the period in $z$ does not contribute. For the topological contribution, however, higher orders do matter. Fortunately, there is an easy way to include them: In the following, $\alpha_0$ is understood to be evaluated at $p+\delta p$ \eqref{delp}, which is the correct elliptic modulus at $\langle B\rangle$. This creates contributions to {\it all} orders in $\epsilon$. These orders will eventually be absorbed in the CSL free energy, to which we need to compare the free energy of our 3D crystal. With this interpretation of $\alpha_0$ in mind and using the equations of motion for $\alpha_0$ (\ref{alpha0EoMEx}) and $\delta\alpha$ (\ref{deltaalphaEoMEx}) we derive
\be
\langle(\nabla\alpha)^2 -2m_\pi^2(\cos\alpha-1)\rangle = \langle(\nabla\alpha_0)^2 -2m_\pi^2(\cos\alpha_0-1)\rangle +\frac{\hat{\omega}_0m_\pi^2}{f_\pi^4}\sum_{\vec{k}} \hat{\omega}(\vec{k}_\perp)\hat{t}(k_z)\delta\hat{\alpha}(\vec{k}) \, .
\ee
Next, and straightforwardly,
\be
\lambda_*\langle|\varphi_0|^4\rangle = e^2\kappa^2\sum_{\vec{k}}\hat{\omega}^2(\vec{k}_\perp)\hat{s}^2(k_z) \, .
\ee
Finally, the WZW contribution, the last term in the free energy 
(\ref{FreeEnergyEx}), is a pure boundary term and does not receive any corrections beyond the term already present in the CSL free energy. Putting everything together yields 
\bea
F&=& F_0 - \frac{e^2\hat{\omega}_0^2}{2}[\beta(2\kappa^2-1)+1]-e^2\sum_{\vec{k}_\perp}\sum_{\ell=1}^\infty \left(2\kappa^2-\frac{k_\perp^2}{k^2}\right) \hat{\omega}^2(\vec{k}_\perp)\hat{s}^2(\ell) \non[2ex]
&&+\,\frac{\hat{\omega}_0m_\pi^2}{2f_\pi^2}\sum_{\vec{k}} \hat{\omega}(\vec{k}_\perp)\hat{t}(k_z)\delta\hat{\alpha}(\vec{k}) \non[2ex]
&=& F_0+\Delta f (\langle B\rangle-B_{c2})^2 \, , 
\eea
where $F_0$ is the CSL free energy density from Eq.\ (\ref{F0}) evaluated at $B=\langle B\rangle$ (i.e., with elliptic modulus $p+\delta p$) and where we have denoted the dimensionless free energy difference  between the pion crystal and CSL by 
\be \label{df}
\Delta f \equiv -\frac{{\cal G}^2}{2}
\frac{1}{\beta(2\kappa^2-1)+1+2{\cal H}_1  - 2{\cal H}_2} \, .
\ee
To derive this result we have used Eq.\ (\ref{EOM0}).
The 3D pion crystal is preferred over CSL if $\Delta f <0$. We may first determine the  energetically favored transverse lattice structure, which is parameterized by $a$. To this end, we plot $\Delta f$ as a function of $a$ for fixed values of $p$ and $m_\pi$, see left panel of Fig.\ \ref{fig:df}. For the values of $p$ chosen in this plot, there are two stationary points: the unstable point $a=1$ (square lattice) and the minimum $a=1/\sqrt{3}\simeq 0.577$ (hexagonal lattice). This is exactly the same observation as in the ordinary type-II superconductor. The $a$ dependence of the functions ${\cal H}_1$ and ${\cal H}_2$ does not change this conclusion; both functions themselves have the same two stationary points.

A more striking result is obtained by exploring the $b$ dependence of $\Delta f$. In the right panel of Fig.\ \ref{fig:df} we plot $\Delta f$ as a function of the critical magnetic field for three different pion masses and fixed $a=1/\sqrt{3}$. Let us first discuss the result for the physical pion mass (black curves). For large critical magnetic fields we find $\Delta f<0$, which indicates that the 3D pion crystal is favored over CSL. Then, as we go to smaller $eB_{c2}$, the denominator in Eq.\ (\ref{df}) changes sign. This can be interpreted as a type-I/type-II transition point (keeping in mind that, as discussed in Sec.\ \ref{sec:results}, our system is not expected to behave like an ordinary type-I superconductor). As we approach this transition point we also observe that $\Delta f$ diverges, indicating that our expansion breaks down. Further decreasing $eB_{c2}$, we see that there is a point where $\Delta f=0$. At this node, ${\cal G}$ changes sign, as discussed in Sec.\ \ref{sec:eomega0}. Since ${\cal G}>0$ for all $eB_{c2}$ larger than at this node, the solution where $\Delta f<0$ is indeed a physical solution for $\langle B\rangle >B_{c2}$.  Finally, the curve ends at the dashed line, which indicates the point $e B_{c2}=3m_\pi^2$. This is the point where the CSL instability curve ends and CSL has turned into a single domain wall. We conclude that the 3D pion lattice is a solution to the instability for sufficiently large $e B_{c2}$, i.e., only for configurations where the picture of well-separated domain walls does {\it not} apply. Had we found $\Delta f<0$ for $eB_{c2}$ {\it smaller} than at the point where ${\cal G}$ changes sign, that would have indicated a physically valid, energetically preferred solution for $\langle B\rangle <B_{c2}$, just below the lower branch of the CSL instability curve and including a transverse lattice confined in a domain wall. Since this is not the case, the single domain wall does not turn into a charged pion lattice, at least not continuously\footnote{It is possible to compute the Abrikosov lattice confined in a single domain wall more directly, by Fourier transforming only the transverse directions and solving the resulting differential equations for the longitudinal direction in position space (which is possible  analytically with the help of hypergeometric functions). We have done this calculation as a consistency check and found agreement with the $p\to 1$ limit of the calculation presented here.}. 

\begin{figure} [t]
\begin{center}
\includegraphics[width=0.5\textwidth]{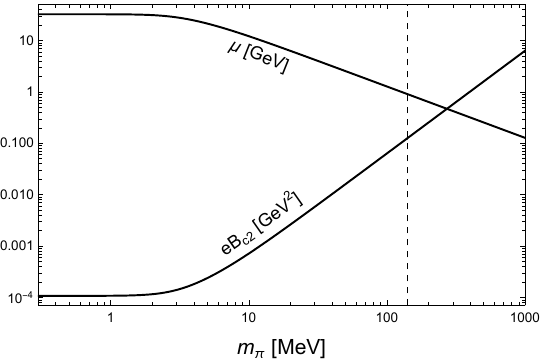}
\caption{Chemical potential and magnetic field of the type-I/type-II transition point as a function of the pion mass. The dashed vertical line indicates the physical point, where $(\mu,eB_{c2})\simeq (907.7 \, {\rm MeV}, 0.1234 \, {\rm GeV}^2)$, as shown in the phase diagram in Fig.\ \ref{fig:pd}.}
\label{fig:mueBmpi}
\end{center}
\end{figure} 

The type-I/type-II transition point moves towards smaller critical magnetic fields (and thus to larger $\mu$) as we decrease the pion mass. This is demonstrated by the red and blue curves in the right panel. In particular, the blue curve is the chiral limit from Ref.\ \cite{Evans:2022hwr}. The dependence of the type-I/type-II transition point on the pion mass is shown in Fig.\ \ref{fig:mueBmpi}. The two curves give the coordinates of the point $(\mu,e\langle B\rangle)$, shown for the physical pion mass as a star in the phase diagram of Fig.\  \ref{fig:pd}. We see that this point approaches a finite limit for vanishing pion mass (which sits at a chemical potential far beyond the validity of chiral perturbation theory) and that for large pion masses it moves up the CSL instability line without any finite boundary.

\begin{acknowledgments}
We would like to thank Tom\'a\v{s} Brauner for useful discussions and comments. G.W.E.\ is supported by the National Science and Technology Council (Taiwan) under Grant No. MOST 110-2112-M-001-070-MY3. G.W.E.\ also acknowledges support by the Deutsche Forschungsgemeinschaft (DFG, German Research Foundation) through the CRC-TR 211 `Strong-interaction matter under extreme conditions'– project number 315477589 – TRR 211, and would like to thank Prof.\ Dirk Rischke and Goethe University Frankfurt's High Energy Physics group for their warm welcome and hospitality during their research visit. 
\end{acknowledgments}

\appendix

\section{Fourier transform of the 2D Abrikosov lattice}
\label{app:fourier}

In this appendix, we derive the result (\ref{omeganm}), i.e., the Fourier transform $\hat{\omega}$ of $\omega\equiv |\phi_0|^2$. 
To simplify the notation in this appendix, let us work with dimensionless quantities $x$, $y$, $q$, obtained from their dimensionful counterparts by rescaling with the coherence length $\xi$, 
\be \label{dimless}
x\to \xi x  \, , \qquad y\to \xi y  \,, \qquad q\to \frac{q}{\xi} \, . 
\ee
Then, with the definition (\ref{omegadef}) and Eq.\ (\ref{phi0}) we have 
\be \label{omxy}
\omega(x,y) = \sum_{n_1,m_1} C_{n_1} C_{m_1}^* e^{i(n_1-m_1)qy} e^{-\left[x^2-(n_1+m_1)qx +\frac{n_1^2+m_1^2}{2}q^2\right]} \, .
\ee
In this appendix, all sums are from $-\infty$ to $+\infty$. We insert this expression into the definition of the inverse Fourier transform (\ref{FourierInv}). The $y$ integral gives a Kronecker delta, which we use to eliminate the sum over $m_1$. With Eq.\ (\ref{LxLy}) and after rewriting the exponential conveniently, we obtain 
\be \label{omnmapp}
\hat{\omega}(n,m) = \frac{e^{-\frac{n^2q^2}{4}}e^{-\frac{\pi^2(n+2m)^2}{4q^2}}}{2q}\int_0^{2q} dx \sum_{n_1}C_{n_1}C_{n_1-n}^* \frac{ e^{-i\frac{\pi}{2}(n+2m)(2n_1-n)}}{e^{\left[x-\left(n_1-\frac{n}{2}\right)q+i \frac{\pi}{q} \left(\frac{n}{2}+m\right)\right]^2}}\, .
\ee
Absorbing for now the irrelevant details in the function $F(n_1)$, we compute by splitting the sum into even and odd contributions, 
\bea\allowdisplaybreaks
\sum_{n_1}C_{n_1}C_{n_1-n}^* F(n_1) &=& \sum_{n_1}C_{2n_1}C_{2n_1-n}^* F(2n_1)+\sum_{n_1}C_{2n_1+1}C_{2n_1+1-n}^* F(2n_1+1) \non[1ex]
&=& \left\{\begin{array}{cc} 
|C|^2\sum_{n_1}F(n_1) & \mbox{for even $n$} \\[2ex] 
\mp i|C|^2\sum_{n_1}(-1)^{n_1}F(n_1) & \mbox{for odd $n$} \end{array}\right. \non[2ex] 
&=& e^{\mp i\frac{\pi}{2}n^2}|C|^2\sum_{n_1}(-1)^{n_1n}F(n_1) \, ,\label{sumFnn1}
\eea
where, in the second step, we have used the form of the coefficients $C_n$ given in Eq.\ (\ref{Cn}). Note that $e^{\mp i\frac{\pi}{2}n^2}$ is 1 for even $n$ and $\mp i$ for odd $n$. Inserting Eq.\ (\ref{sumFnn1}) into Eq.\ (\ref{omnmapp}), rewriting $e^{-i\frac{\pi}{2}(n+2m)(2n_1-n)} =  (-1)^{n_1n} (-1)^{nm}e^{i\frac{\pi}{2}n^2}$, and using (for the lower sign) $e^{i\pi n^2} = (-1)^n$, we find 
\bea
\hat{\omega}(n,m) &=& |C|^2 (\pm 1)^n (-1)^{nm} \frac{e^{-\frac{n^2q^2}{4}}e^{-\frac{\pi^2(n+2m)^2}{4q^2}}}{2q}\int_0^{2q} dx \sum_{n_1}  e^{-\left[x-\left(n_1-\frac{n}{2}\right)q+i \frac{\pi}{q} \left(\frac{n}{2}+m\right)\right]^2}  \, . \;\;\;\; \;\;\;\label{omnm1}
\eea
The integral is  
\be
\int_0^{2q} dx \sum_{n_1}  e^{-\left[x-\left(n_1-\frac{n}{2}\right)q+i \frac{\pi}{q} \left(\frac{n}{2}+m\right)\right]^2} = 2\sqrt{\pi}  \, ,
\ee
where odd and even $n_1$ both give a Gaussian integral over the domain $[-\infty,\infty]$, hence the factor 2. Inserting this into Eq.\ (\ref{omnm1}) and using Eq.\ (\ref{adef}) gives the result (\ref{omeganm}).

\bibliographystyle{JHEP}
\bibliography{references2}

\end{document}